\documentclass[11pt,a4paper]{article}
\usepackage{jheppub}
\usepackage{amsmath,xfrac,xcolor,amssymb,float,cancel}

\def\susy{{\cancel{\text{susy}}}}
\def\Re{\text{Re}\,}
\def\Im{\text{Im}\,}

\title{Linking the Supersymmetric Standard Model to the Cosmological Constant}

\author[a,b]{Yu-Cheng Qiu}
\author[a,b,c]{and S.-H. Henry Tye}

\affiliation[a]{Jockey Club Institute for Advanced Study,\\
Hong Kong University of Science and Technology, Hong Kong S.A.R., China}
\affiliation[b]{Department of Physics,\\
Hong Kong University of Science and Technology, Hong Kong S.A.R., China}
\affiliation[c]{Department of Physics,\\
Cornell University, Ithaca, NY 14853, USA}

\emailAdd{yqiuai@connect.ust.hk}
\emailAdd{iastye@ust.hk}

\abstract{
String theory has no parameter except the string scale $M_S$, so the Planck scale $M_\text{Pl}$, the supersymmetry-breaking scale $m_\susy$, the electroweak scale $m_\text{EW}$ as well as the vacuum energy density (cosmological constant) $\Lambda$ are to be determined dynamically at any local minimum solution in the string theory landscape. Here we consider a model that links the supersymmetric electroweak phenomenology (bottom up) to the string theory motivated flux compactification approach (top down). In this model, supersymmetry is broken by a combination of the racetrack K\"ahler uplift mechanism, which naturally allows an exponentially small positive $\Lambda$ in a local minimum, and the anti-D3-brane in the KKLT scenario. In the absence of the Higgs doublets from the supersymmetric standard model, one has either a small $\Lambda$ or a big enough $m_\susy$, but not both. The introduction of the Higgs fields (with their soft terms) allows a small $\Lambda$ and a big enough $m_\susy$ simultaneously. Since an exponentially small $\Lambda$ is statistically preferred (as the properly normalized probability distribution $P(\Lambda)$ diverges at $\Lambda=0^{+}$), identifying the observed $\Lambda_{\rm obs}$ to the median value $\Lambda_{50\%}$ yields $m_{\rm EW} \sim 100$ GeV. We also find that the warped anti-D3-brane tension has a SUSY-breaking scale $M_\susy\sim100m_{\rm EW}$ while the SUSY-breaking scale that directly correlates with the Higgs fields in the visible sector is $m_\susy\simeq m_{\rm EW}$.

}


\arxivnumber{2006.16620}

\begin{document}

\maketitle

\section{Introduction}\label{sec:intro}

One of the guiding principles in physics is naturalness, that is, why a particular energy/mass scale emerges without fine-tuning. Consider the four-dimensional effective action
\begin{equation}
	\label{Action0}
	\mathcal{S}=\int {\rm d}^4x \sqrt{-g}\left[-\Lambda +\frac{M_\text{Pl}^2}{2}R - \frac{m_h^2}{2}h^2 + \cdots\right]\;,
\end{equation}
which displays some of the most relevant operators that are known to be present in nature. 
Here we encounter the puzzle why the Higgs boson $h$ has mass $m_h =125$ GeV (or the electroweak (EW) scale $m_\text{EW} \simeq 10^2$ GeV) which is much smaller than the (reduced) Planck mass $M_\text{Pl} = 2.4 \times 10^{18}$ GeV, 
\begin{equation}
m^2_\text{EW} \simeq 10^{-32}M_\text{Pl}^2\;,
\label{mEWP}
\end{equation}
as naive radiative correction from quantum loop effects tend to contribute an order of $M_\text{Pl}^2$ to $m_h^2$. This puzzle is known as the mass hierarchy problem. It motivates the study of supersymmetry (SUSY), and supersymmetric standard model (SSM) phenomenology has been the mainstream theoretical investigation beyond the standard model in the past decades \{cf.\cite{Martin:1997ns,Weinberg:2000cr,Baer:2006rs}\}. If SUSY is present, with the SUSY-breaking scale in the visible sector of order of $m_\susy \sim m_\text{EW}$, the radiative corrections to $m_h$ would be of order $m_{\rm EW}$. However, this mechanism does not explain why the tree-level Higgs mass is not of order $M_\text{Pl}$. So this approach is known to be technically natural only, not natural; that is, we still have the so called $\mu$ problem.

More seriously, the most relevant operator (or term) in $\mathcal{S}$\eqref{Action0}, namely the observed vacuum energy density, or the cosmological constant $\Lambda$, where\footnote{If the dark energy density has another origin, then $\Lambda$ has to be further fine-tuned to a smaller value.}
\begin{equation}
	\label{Lambda}
	\Lambda_\text{obs} \simeq 10^{-120}M_\text{Pl}^4\;,
\end{equation}
clearly is a bigger puzzle. Here again, radiative corrections from the standard model to $\Lambda$ will be at least as big as $m_\text{EW}$ scale, which is many orders of magnitude too big compared to the observed $\Lambda_{\rm obs}$ (this is known as the radiative instability problem). Since $\Lambda$ is a free parameter within the quantum field theory (QFT) framework, there is no chance to understand its small value naturally without going beyond QFT. Fortunately, $\Lambda$ is calculable in string theory. 

String theory has no parameter ($c=\hbar=1$) except the string scale $M_S$. So both $M_\text{Pl}$ and $\Lambda$ can be dynamically determined in terms of $M_S$ for any (meta-)stable vacuum and we can express $\Lambda$ in terms of $M_\text{Pl}$, so there is a chance to address this puzzle in string  theory. To see how an exponentially small $\Lambda$ can emerge naturally, we shall focus on the brane world scenario\footnote{The picture is not dissimilar to the coronavirus, where the Calabi-Yau bulk (body) has multiple warped throats (spikes) attached to it.} where the ten-dimensional spacetime in Type IIB string theory is flux compactified with a Calabi-Yau orientifold to four-dimensional spacetime. The standard model particles are open string modes inside a stack of D3-branes (plus D7-branes wrapping a 4-cycle) sitting in a warped throat. The graviton, the dilaton $S$, the complex structure (shape) moduli $U_i$ and  the K\"ahler (size) moduli $T_j$ are closed string modes. Scanning over the discrete flux values \cite{Bousso:2000xa} and the geometry generates the string landscape. So scales we obtain are statistical in nature.

\begin{figure}
	\centering
	\includegraphics{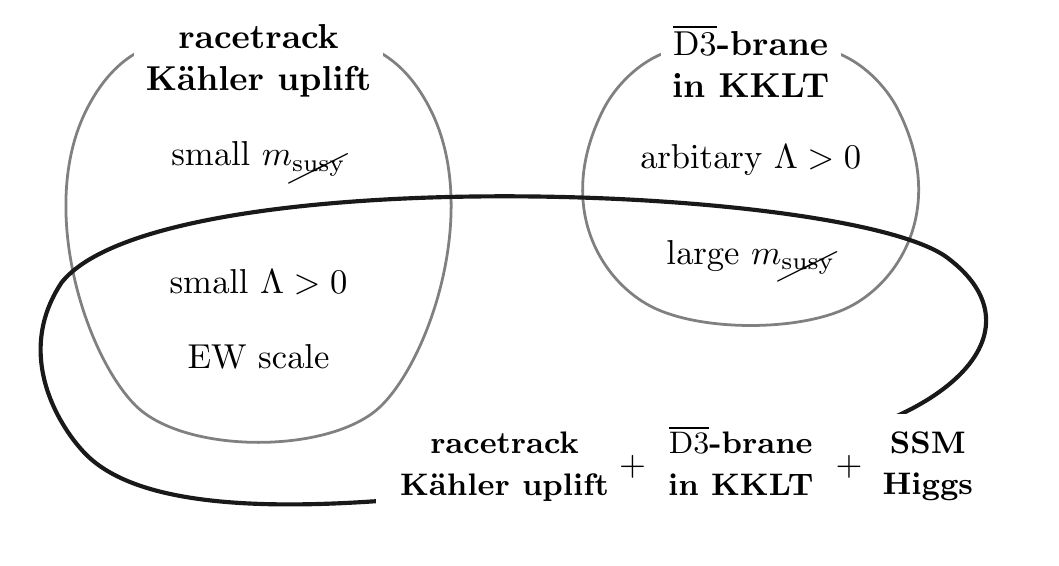}
	\caption{Relations among the 3 pillars of the model : the KKLT scenario with $\overline{\rm D3}$-branes, the racetrack  K\"ahler uplift model and the supersymmetric standard model with the two electroweak Higgs doublets. In the absence of the soft SUSY-breaking terms for the Higgs fields, one can have an exponentially small $\Lambda>0$ or a large enough SUSY-breaking scale $m_{\susy} \sim m_{\rm EW}$, but not both.}
	\label{fig:model}
\end{figure}

One may view our approach as the KKLT scenario \cite{Kachru:2003aw} \{cf.\cite{Douglas:2006es,Becker:2007zj,Ibanez:2012zz}\}, implemented with the racetrack K\"ahler uplift (RKU) model \cite{Sumitomo:2013vla} (which combines the racetrack \cite{Krasnikov:1987jj,Taylor:1990wr,Denef:2004dm} with the K\"ahler uplift (KU) model \cite{Balasubramanian:2004uy,Westphal:2006tn,Rummel:2011cd,deAlwis:2011dp,Sumitomo:2012vx,Louis:2012nb}) and linking them to SSM of strong and electroweak interactions. That is, we are considering the four-dimensional low energy effective potential for the patch of landscape that includes the SSM (assuming it is a solution in string theory). Interesting properties reveal themselves when we put them together with the SSM phenomenology.  It turns out that all three ingredients are needed for an exponentially small $\Lambda$ and  $m_\susy \simeq m_\text{EW} \sim 100$ GeV, as shown in figure \ref{fig:model}. Some of the known steps in this program can be found in the literature. 
Let us sketch the overall picture here, step by step. 

$\bullet$ It is well known that, in the weak string coupling $g_s \simeq 1/{\Re(S)}$ region, there is no de Sitter (dS) vacuum as the vacuum expectation value (vev) of the dilaton  $\Re(S) \to \infty$ \cite{Dine:1985he}. The reasoning has been extended to other moduli in the context of the swampland and distance conjecture (e.g.cf.\cite{Ooguri:2018wrx}), including the K\"ahler modulus $T=t + i \tau  \propto x+iy$, the scalar mode that determines the volume of the compactified Calabi-Yau manifold. In the asymptotic regime $x \to \infty$, the perturbative $V_p(x) \to 0$, as the compactified volume approaches infinity, i.e., the six-dimensional manifold decompactifies, and there is no dS vacuum solution.  

Fortunately, a non-perturbative term (from e.g., gaugino condensation) of the form $e^{-x}$ can be introduced \cite{Novikov:1983ee,Novikov:1983ek,Ferrara:1982qs,Dine:1985rz,Derendinger:1985kk,Shifman:1987ia,Gorlich:2004qm}, where now we have something like 
$V(x) = Ae^{-x} + V_p(x)$,
so a local minimum at large but finite $x$ may exist (see appendix \ref{appendix:review}). Here we define naturalness in the following way: a small $\Lambda$ is natural if $\Lambda \sim e^{-x}>0$ in a large $x$ (e.g., we have in mind $x \sim \mathcal{O}\left(100\right)$) solution. If there is no large $x$ solution, then one has to fine-tune some parameters in the model to obtain a small $\Lambda$, which is not natural. Here, in this simple model, where the dilaton and all complex structure moduli have been stabilized already \cite{Giddings:2001yu}, $\Lambda \propto e^{-x}$  can be exponentially small. However, the solution yields an anti-de Sitter (AdS) supersymmetric vacuum with $\Lambda <0$. 

$\bullet$ There are two well studied mechanisms in string theory to lift the AdS vacuum to a meta-stable dS vacuum, hence breaking SUSY :\\
(1)  the introduction of an anti-D3 ($\overline{\rm D3}$-)brane in the KKLT model \cite{Kachru:2003aw}, which explicitly breaks SUSY; \\
(2)  a string theory $\alpha'^3$-correction in the K\"ahler potential in KU model \cite{Balasubramanian:2004uy,Westphal:2006tn,Rummel:2011cd,deAlwis:2011dp,Sumitomo:2012vx}, which provides a $F$-term spontaneous SUSY breaking. \\
Both scenarios can yield a dS (or AdS) vacuum, but the value $\Lambda$ is essentially unrestricted. Without fine-tuning,
its magnitude is typically much too big to yield $\Lambda_{\rm obs}$ (\ref{Lambda}). Although the KKLT model is relatively unconstrained, the KU model turns out to have a rather non-trivial constraint.

$\bullet$  A dS solution in the KU model restricts $x<3.12$ \cite{Rummel:2011cd}, while only an AdS solution is available for large $x$ (see appendix \ref{appendix:review}). So one has to fine-tune the available parameters in the model to obtain an exponentially small $\Lambda$. 

This bound on $x$ is relaxed if we introduce the racetrack to the KU model (see appendix \ref{appendix:racetrack}), where two non-perturbative terms compete with each other (hence the name racetrack). Racetrack is very natural in string theory \cite{Krasnikov:1987jj,Taylor:1990wr,Denef:2004dm} and it does wonders here. In the RKU model \cite{Sumitomo:2013vla}, stabilizing both $x$ and $y$ at a local minimum put tight constraints on the existence of a dS vacuum: $\Lambda$ is bounded by 
\begin{equation}
\label{ulbound}
0 <\Lambda_{\rm min} \le \Lambda \le \Lambda_{\rm max}\;,
 \end{equation}
where both $\Lambda_{\rm min}$ and $\Lambda_{\rm max}$ go like $e^{-2x}$ for large $x$, and $\Lambda_{\rm min}/\Lambda_{\rm max} \to 1$ as $x \to \infty$. This implies that the range of possible $\Lambda$ solution is very small for large $x$, putting tight constraints on the input parameters for any solution to exist. An examination of the statistical distribution of $\Lambda$ implies that $\Lambda$ can be naturally exponentially small. In short, racetrack trades the (undesirable) bound on $x$ in the KU model for the (powerful) bounds on $\Lambda$ in the RKU model.

Before describing more results, let us recall our approach \cite{Sumitomo:2013vla}, as the model does have a number of parameters  (like those in eq.(\ref{VT1}) and those in eq.(\ref{3amigos})). For some choices of their values, there is no local (meta-stable) minimum.
For some restricted choices of parameters that allow a minimum solution, the resulting $\Lambda$ is a function of those parameters, which in turn are functions of the discrete flux values $F^i$ that are present in the flux compactification \cite{Bousso:2000xa}. The string theory landscape is generated by scanning over the ``dense discretuum''  of all the discrete flux values. To quantify the properties, we determine the probability distribution $P(\Lambda)$ of the $\Lambda$ value by sweeping through all flux parameter values that yield a local minimum at some finite $x$ value.

 In supergravity (SUGRA), we have the $F$-term potential (where $M_\text{Pl}=1$),
\begin{equation}
\label{VT1}
V(T) = e^K\left(|DW|^2 - 3 |W|^2\right)\;, \quad \quad W=  Ae^{-x} + Be^{-\beta x} +\mathcal{W}\;,
\end{equation}
where the K\"ahler potential $K=-2\ln{\left(\left(T + \bar{T}\right)^{3/2}+\xi/2\right)} + \cdots$ contains the string correction term $\xi>0$ which does the uplift, and $A$, $B$, $\beta$ and $\mathcal{W}$ are input parameters independent of $x$ in this approximation. So
\begin{equation}
V(x) \sim e^{-2x} + e^{-x} \mathcal{W} + |\mathcal{W}|^2+ \cdots\;,
\label{3termcase}
\end{equation}
where some coefficients and $x$-dependence are suppressed. The bounds (\ref{ulbound}) strongly limit the existence of any dS vacuum solution, thus imposing tight constraints on the parameters of the model, including $\mathcal{W}$. At the minimum where these terms balance each other, $\mathcal{W} \sim e^{-x}$ and $\Lambda \sim e^{-2x}$. (As the $x$-independent $\mathcal{W}$ takes a wide range of values in the landscape, a minimum solution exists only for such a small $\mathcal{W}$.)
With a naturally exponentially small $\Lambda$, simple dimensional arguments suggests
\begin{equation}
\Lambda_\text{obs} \sim {\mathcal{W}}^2 \sim 10^{-120} \; \to\; \mathcal{W}\sim {\bf m}^3 \sim 10^{-60} \;\to\; {\bf m} \sim 10^{-20}
\end{equation}
where a new scale ${\bf m}$ automatically emerges. In ref.\cite{Andriolo:2018dee}, including factors coming from the RKU model \cite{Sumitomo:2013vla}, ${\bf m} \sim 10^{2}$ GeV emerges without making any reference to SSM. If the electroweak contribution (the two Higgs doublet term $\mu h_1 h_2$)
inside $\mathcal{W}$ is not negligible, one is led to identify ${\bf m} \simeq m_{\rm EW}$.

$\bullet$  Unfortunately, the above RKU model yields a negligibly small gravitino mass $m_{3/2}$, which measures the SUSY-breaking scale $m_{\susy}$. Clearly this $m^2_{3/2} \sim |\Lambda|/M_{\rm Pl}^2$ is not suitable for SSM phenomenology \cite{Sumitomo:2013vla,Tye:2016jzi}. To have a large enough $m_\susy$ to fit phenomenology, we introduce the other known mechanism, namely the $\overline{\rm D3}$-brane in the KKLT scenario. However, $\overline{\rm D3}$-brane alone does not yield an exponentially small $\Lambda$ without fine-tuning. This leads us to input both SUSY-breaking mechanisms, RKU and $\overline{\rm D3}$-brane. Even then, the dynamics does not work out (i.e., the warped $\overline{\rm D3}$-brane tension is forced to a negligibly small value) unless we introduce the Higgs fields as well. This is the main point of this paper. Let us extend $V(T)$ (\ref{VT1}) to 
\begin{equation}
\label{VT3}
V(T) = e^K\left(|DW|^2 - 3 |W|^2\right) + V_3 + D_h + S_h \;,
\end{equation}
where $V_3$ is the potential from a small stack of $p$ $\overline{\rm D3}$-branes with warped brane tension, $D_h$ is $D$-term for Higgs fields in SSM and $S_h$ contains the soft terms (i.e., terms that preserve technical naturalness \cite{Girardello:1981wz}) in the Higgs sector. Recalling that the Higgs potential at the minimum after spontaneous symmetry breaking (SSB) $|h_i^{0}|=v_i$ reduces to (not including terms suppressed by powers of $M_{\rm Pl}$),
\begin{align}
\label{Higgs1}
	\left.V_{h}\right|_{\rm min}
	&=|\mu|^2\left(v_1^2 +v_2^2\right) +S_h +D_h <0\;, \nonumber \\
	S_h&=m_1^2v_1^2+m_2^2v_2^2-2bv_1v_2\;,\quad D_h=\frac{1}{8}\left(g^2+g'^2\right)\left(v_1^2-v_2^2\right)^2\;.
\end{align}
where the $\mu$ term comes from the $\mu h_1h_2$ term inside $\mathcal{W}$.
Here $\left.V_{h}\right|_{\rm min}$, a function of vev's $v_1$ and $v_2$, is negative in general as $V_h(v_i=0)=0$. (Quantum corrections can be included in $V_h$ without changing our analysis.) This is the phenomenological input in our model, where the scale of the parameters $\mu$, $m_i$, $b$ and the gauge couplings are {\it A priori} undetermined. Our goal is to see how the scale $m_{\rm EW} \simeq |\mu v_1v_2|^{1/3} \simeq 100$ GeV emerges. Finding that $\Re(S)=g_s^{-1} \gtrsim 1$, it follows that  $v_i \sim m_{\rm EW}$ implies that $|m_i|$ and $b$ also are of order $m_{\rm EW}$.

We write those three terms in $V(T)$ (\ref{VT3}) together as
\begin{align}
\label{3amigos}
V_3+D_h+S_h &= \frac{\mathcal{D}}{(T+\bar{T})^n}=\frac{p\tilde{T}_3+ \hat{D}_h + \hat{S}_h}{(T+\bar{T})^2}\;, 
\end{align}
where $\tilde{T}_3$ is the warped $\overline{\rm D3}$-brane tension that breaks SUSY explicitly in the landscape, and we introduce the scale of SUSY-breaking in the landscape to be $M_{\susy}$ where  $M^4_{\susy} \simeq \tilde{T}_3$. The hats in $\hat{D}_h$ and $\hat{S}_h$ indicate that the values in them have to be rescaled by the $(T+\bar{T})$ factors to match those in $V_h$ (\ref{Higgs1}). Here the power of $(T+\bar{T})$ is set at $n=2$ \cite{Kachru:2003sx,Kachru:2019dvo} (see appendix \ref{appendix:barD3}). 

\begin{figure}
	\centering
	\includegraphics[width=7cm]{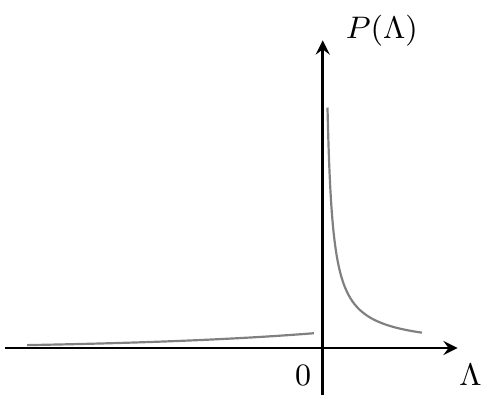}
	\caption{Sketch (not to scale) of the properly normalized probability distribution $P(\Lambda)$ for the cosmological constant $\Lambda$ as expected from the model, where $P(\Lambda)$ is power-divergent as $\Lambda \to 0^+$. This implies that an exponentially small positive $\Lambda$ is statistically preferred in the patch of string theory landscape described by the model.} 
	\label{fig:proL}
\end{figure}

It is crucial that these sectors must couple to each other via closed string interactions, as required in string theory. Here  the K\"ahler modulus $T$ plays that role. Otherwise, any term in $V(T)$ (\ref{VT3}) that is independent of $T$ will simply shift $\Lambda$ by an amount of order of $m_{\rm EW}^4$ or $M_{\susy}^4$, which is unacceptably large for a naturally small $\Lambda$. Here $D_h>0$, $V_3>0$ and crucially, $S_h<0$, which is needed for SSB to happen. Since the soft terms are present when SUSY is broken,  $S_h \to 0$ as $\tilde{T}_3 \to 0$ (the $\xi$-term breaking of SUSY is negligibly small), it is reasonable to assume that $\mathcal{D}\equiv\hat{D}_h + p \tilde{T}_3 +\hat{S}_h > 0 $. Now, the bound $\Lambda_{\rm min} < \Lambda_{\rm max}$ (\ref{ulbound}), which is required for a local minimum to exist, forces $\mathcal{D} \to 0$. (In the absence of the Higgs terms, this means $\tilde{T}_3 \to 0$.) In the resulting meta-stable dS solution, \\
(1) $\Lambda \propto e^{-2x} >0$ is naturally exponentially small, since $P(\Lambda)$ diverges as $\Lambda \to 0^+$;\\
(2) the EW scale $m_{\rm EW}$ emerges from the observed $\Lambda_{\rm obs}$, 
and \\
(3) the SUSY-breaking scale is about or above the EW scale.\\
In the unlikely case where $\mathcal{D} < 0$, we may obtain AdS solutions, but $P(\Lambda)$ is relatively smooth for $\Lambda <0$, as illustrated in figure \ref{fig:proL}.

Here, in our model, we assume all light scalar fields $\phi_i$ except the single K\"ahler modulus $T$ have already been stabilized, so the potential $V\left(\phi_{0j}(F^i), \mathcal{A}_k(F^i), T\right)$ is a function of $T$ only and we have $\Lambda (F^i)$ at a minimum. As we scan over all discrete flux values (in practice, we scan over random values of all the parameters $\mathcal{A}_k$ (and the moduli and dilaton $\phi_{0j}(F^i)$) that yields a vacuum solution to obtain the properly normalized $P(\Lambda)$.  We find that our model yields $P(\Lambda) \propto \Lambda^{-1+1/2N}$ ($N$ is the rank of the $D7$ gauge symmetry)  that diverges at $\Lambda=0^{+}$. That is, not only that $\Lambda \propto e^{-2x}$ at large $x$ is a solution, it is statistically preferred; so a typical $\Lambda$ is positive and naturally exponentially small (see figure \ref{fig:proL}).  Choosing a $P(\Lambda)$ (i.e., choosing $N \sim 200$) whose median $\Lambda_{50\%}$ matches the observed $\Lambda_{\rm obs}$ (which requires a small $k$), we obtain (with $|\mu v_1v_2| \simeq m^3_{\rm EW}$), with roughly a factor of two uncertainty,
\begin{equation}
\label{result}
\Lambda_\text{50\%}=\Lambda_\text{obs} \; \to \; m_\text{EW} \simeq m_{\susy} \sim 100 \, {\rm GeV}\;, \quad   M_\susy \sim  100 \, m_\text{EW}
\end{equation}
where $ M_\susy$ is the SUSY-breaking scale in the landscape while $m_{\susy}$ is the SUSY-breaking scale directly correlated with the EW scale in the visible sector. We note that the region with a dS solution seems to be a very small region in the landscape, much like an oasis in a desert. 

Referring back to figure \ref{fig:model}, let us summarize the key observations here: \\
$\bullet$ The KKLT model with an $\overline{\rm D3}$-brane breaks SUSY. It can yield a AdS or dS vacuum, where values of both $M_{\susy}$ and $\Lambda$ are hardly constrained. \\
$\bullet$ The KU model has no dS solution except for a relatively small $x<3.12$, so $\Lambda$ is not naturally small. The RKU model allows a dS solution where $x$ can be large, so $\Lambda \propto e^{-2x}$ can be small. Divergence of $P(\Lambda)$ at $\Lambda=0^+$ implies that an exponentially small positive $\Lambda$ is preferred.\\
$\bullet$ Introducing both SUSY-breaking mechanisms in the racetrack model allows an exponentially small $\Lambda$ but the upper bound on $\tilde{T}_3 =M^4_{\susy}$ only allows an unacceptably small $M_{\susy}$. The introduction of Higgs doublets coverts an upper bound on $\tilde{T}_3$ to an upper bound on $\mathcal{D}$ (\ref{3amigos}), freeing $\tilde{T}_3$ to be large enough for electroweak phenomenology. That is, the presence of the electroweak Higgs doublets raises $M_{\susy}$ to a scale higher than or comparable to $m_{\rm EW}$. \\
$\bullet$ At first sight, it seems that we have to fine-tune the warped $\overline{\rm D3}$-brane tension just to obtain a solution, where $m_{\susy} \simeq m_{\rm EW} \sim 10^{-16}M_\text{Pl}$. We believe it is against naturalness to have two different throats with the same warp factor. So we view this as strong evidence that both SSM and the $\overline{\rm D3}$-branes live at the bottom of the same warped throat in the compactified manifold. If SSM comes from a stack of $D3$-branes, it is important to check whether the stack of $D3$-branes and an $\overline{\rm D3}$-brane can co-exist (separately) at the bottom of a deformed and resolved confifold.  On the other hand, it may be more economical (and so more attractive) if the standard model particles come from the same stack of $\overline{\rm D3}$-branes that provide the uplift to dS space \cite{GarciadelMoral:2017vnz,Cribiori:2019hod,Parameswaran:2020ukp}. Here $p \simeq 5$ presumably will not de-stabilize the meta-stable dS vacuum.

Here, the two K\"ahler modes have masses, within a few orders of magnitude, $m_t \simeq m_{\tau} \sim \sqrt{\Lambda/M_{\rm Pl}^2} \sim 10^{-33}$ eV \cite{Tye:2016jzi}. So their energy density contribution in the universe today may still behave like dark energy, making a contribution to $\Lambda$. Only at a later time as the Hubble parameter $H$ reaches a value below the K\"ahler masses when the mis-alignment mechanism happens, converting their contribution in the dark energy density to dark matter density \{cf.\cite{Kolb:1990vq,Marsh:2015xka}\}. This possibility and the possibility that the mis-alignment mechanism is already operating will be studied elsewhere.

It is important to note that we have not found the string theory solution that reproduces the standard model of strong and electroweak interactions. Here, we assume that there is such a solution in string theory, and we try to find the conditions how various scales appear naturally and hopefully provide clues to where the standard model is hiding in the string theory landscape. Even then, we believe that some parameters in the standard model can never be precisely determined; they can only be estimated based on statistical properties of the fluxes involved, as the fermion mass spectrum seems to indicate \cite{Andriolo:2019gcb}. In this sense, 
the model is a minimal scenario that incorporates the key orders-of-magnitude features without fine-tuning. We have not specified how the soft terms in $S_h$ are generated, except that the low $M_{\susy}$ implies that gravity mediated SUSY breaking is not viable here. With SUSY broken explicitly, the soft terms can simply appear for some choice of the flux parameters in the landscape. Clearly, more input is necessary to determine the sparticle spectrum and their interactions. In phenomenology, naturalness (of $\Lambda$, $M_{\susy}$ and $m_{\rm EW}$ scales) can impose strong constraints on some extension/variation of SSM.

Compared to QFT, where we have to fine-tune the parameters in a vacuum solution to fit nature; here instead, we have to tune the flux parameters in the string landscape just to obtain a vacuum solution, in which the mass scales in $\mathcal{S}$ (\ref{Action0}) appear natural. We believe this is a step forward in gaining a new perspective in understanding our universe. In SSM, SUSY breaking leads to the soft terms triggering SSB, while here the soft terms and SSB work in tandem to allow a local minimum solution.

The paper is organised as follows. In section \ref{sec:model}, we present the model, which combines parts that have been studied before, so we shall emphasize the new parts linking the various pieces together. In section \ref{sec:analysis}, we find the dS solution of the model. Scanning over the discrete flux parameters, we obtain the probability distribution $P(\Lambda)$ in section \ref{sec:sta}, which peaks and  diverges at $\Lambda=0^{+}$. Choosing $P(\Lambda)$ in which the median $\Lambda_{50\%}$ matches the observed $\Lambda_{\rm obs}$, we find that the $m_{\rm EW} \simeq m_{\susy} \sim 100$ GeV and $M_{\susy} \sim 100 \, m_{\rm EW}$. This and other properties are presented and explained in section \ref{sec:phe}. Discussions and remarks are contained in section \ref{sec:dis} while summary and conclusion are in section \ref{sec:summary}. 

Some of the details and reviews are contained in the appendices. Although the original choice of the power of $1/(T+\bar{T})$ in $V(T)$ (\ref{3amigos}) is taken to be $n=3$ \cite{Kachru:2003aw}, it was subsequently accepted that $n=2$ \cite{Kachru:2003sx,Kachru:2019dvo}.  We review this issue and extend the reasoning for $n=2$ for the Higgs terms in appendix \ref{appendix:barD3}. Appendices \ref{appendix:review} and \ref{appendix:racetrack} review the solution/analysis of the properties of the various (sub-)models which are parts of the model discussed in the main text, namely, the KKLT model with and without $\overline{\rm D3}$-brane, and the KU model with and without the racetrack. These reviews play the role of a warm up to the more detailed analysis in the text. Comparison between various models also brings out the role each ingredient, namely, the KKLT scenario, the KU uplift and the racetrack, plays in our model. We keep leading order in the analytical calculation while some detailed expressions are contained in appendices \ref{appendix:combined} and \ref{appendix:cal}. Some terms are dropped in our approximation, which is justified in appendix \ref{appendix:drop}. Appendix \ref{appendix:prefactor} discuss the role of the dilaton and the complex structure moduli in the model, which assumes they have already been stabilized and have little impact in the main calculation. Appendix \ref{appendix:stat}
contains some of the details of the statistical analysis of the probability distribution $P(\Lambda)$ of the cosmological constant $\Lambda$.

\section{The Model}\label{sec:model}

Let us present the effective potential in our simplified model. The relation between the string scale $M_S$ and $M_\text{Pl}$ is given in terms of the dimensionless compactified volume $\mathcal{V}$, which is related to the  K\"ahler modulus $T$ given by\footnote{More precisely, $\mathcal{V}= \frac{\sqrt{3}}{2\sqrt{\gamma}} (T + \bar{T})^{3/2}$, where $\gamma$ is the self-intersection number of $T$ in terms of the Poincare-dual 2-cycle volume modulus of the underlying $\mathcal{N}=2$ supersymmetric theory before orientifolding. Here we consistently drop such factors.}
\begin{equation}
\label{V6}
	\mathcal{V}=\left(\frac{M_\text{Pl}}{M_S}\right)^2 = \left(T+\bar{T}\right)^{3/2} \;.
\end{equation}
For our approximation to be valid, $\mathcal{V} \gg 1$, which is the large volume scenario. Although the fundamental scale is $M_S$, it is more convenient to express everything in terms of $M_\text{Pl}$ instead. So everything is dimensionless, which is easy for calculation. Flux compactification of ten-dimensional theory yields the four-dimensional low energy effective potential. With $M_\text{Pl}=1$, we have, in the SUGRA framework, like eq.(\ref{VT3}),
\begin{equation}
 \label{eq:effV}
V(T) = V_F + V_3 + D_h + S_h\;.
\end{equation}

$\bullet$ The first term is the $F$-term potential for the RKU model with SSM Higgs included. It can be obtained from ten-dimensional Type IIB string theory \cite{Kachru:2019dvo,Hamada:2019ack} 
 \begin{equation}
 \label{VF}
V_F(T)=\mathcal{N}(U_i, S)e^K\left(K^{T\bar{T}}D_T W D_{\bar{T}} \bar{W}-3|W|^2\right)\;,
 \end{equation}
 where
 \begin{align}\label{Modele}
	K&=-2\ln{\left((T + \bar{T})^{3/2}+\frac{\xi}{2}\right)}+\hat{h}_i^\dagger \hat{h}_i+\cdots\;,\nonumber \\
	\xi &=-\frac{\zeta(3)}{4\sqrt{2}(2\pi)^3}\chi(\mathcal{M})\left(S+\bar{S}\right)^{3/2}>0\;,\nonumber\\
	W&=\mathcal{W}+W_\text{NP}\;,\quad \mathcal{W}=W_0(U_i,S)+\hat{\mu} \hat{h}_1 \hat{h}_2\;,\nonumber \\
	W_\text{NP}&=Ae^{-aT}+Be^{-bT}\;.
\end{align}
Here we assume that the dilaton $S$ and the complex structure moduli $U_i$ ($i=1,2,...,h^{2,1}$) have been stabilized, so that their contribution inside the K\"ahler potential $K$ has been placed in the normalization factor 
$$\mathcal{N}(U_i, S)=e^{K(U_i,S)}= 1/\left(S+\bar{S}\right) \prod_i \left(U_i+\bar{U}_i\right)\;,$$
and their contribution $W_0(S, U_i)$  inside the superpotential $W$ behaves like a constant. A review can be found in appendix \ref{appendix:prefactor}. The $\xi$-term is an $\alpha'^3$-correction, where the Euler index of the manifold $\mathcal{M}$ is $\chi(\mathcal{M})= 2(h^{1,1}- h^{2,1})$. Since we consider only a single $T$ (i.e., $h^{1,1}=1$), while at least three $U_i$ (i.e., $h^{2,1} \gtrsim 3$), $\chi(\mathcal{M})<0$ and $\xi >0$. For the weak coupling approximation to be valid, $\Re(S) = g_s^{-1} >1$. So typical values of $\xi$ is from $10^{-3}$ to $10^{-2}$. The form of $K(T)$ (\ref{Modele}) yields no-scale SUGRA when $\xi=0$.
Note that, with eq.(\ref{V6}), the typical scale of $V_F(T)$ (\ref{VF})  goes like $V_F(T) \sim M^4_\text{Pl} e^K \sim M^4_\text{Pl}/\mathcal{V}^2 \sim M_S^4$, as expected, since $M_S$ is the fundamental scale in string theory.
 
$\bullet$ In general, a low energy effective potential is valid only in some region of the landscape, not the whole of landscape.  Here we are considering the $V(T)$ (\ref{eq:effV}) describing the patch of landscape that includes the standard model. So we include the SSM Higgs term inside $W$. We put a hat on them because the Higgs fields $h_1$ and $h_2$ in the SSM involves a normalization. Below, we assume SSB has already taken place, so the Higgs fields already have non-zero vevs. We also assume both radiative corrections and renormalization group flow have been carried out and already included here. Other SSM terms in $W$ have zero vev's so they are dropped in $W$ and $V$.

$\bullet$  The non-perturbative terms in $W_\text{NP}$ may emerge in the following way. Besides D3-branes, Type IIB string theory also allows D7-branes. Here, we consider a stack of D7-branes wrapping a 4-cycle inside the compactified manifold, while its remaining three spatial dimensions overlap with the three large spatial dimensions of the stack of D3-branes. The stack of D7-branes is endowed with a pure $\mathcal{N}=1$ super Yang-Mills gauge symmetry. The dimensionful gauge coupling $g_7$ is related to the dimensionless gauge coupling $g_4$, $g_4^{-2} \sim g_7^{-2} \mathcal{V}_4$, where $\mathcal{V}_4$ is the 4-cycle volume. Recall that the six-dimensional volume $\mathcal{V}$  (\ref{V6}) scales with $3/2$ power of $T$, $\mathcal{V}_4$ scales like $T$. In the effective $3+1$ dimensions, gauge interactions become strong and gaugino condensation takes place  \cite{Novikov:1983ee,Novikov:1983ek,Ferrara:1982qs,Dine:1985rz,Derendinger:1985kk,Shifman:1987ia,Berg:2004ek} \{cf.\cite{Baumann:2014nda}\}, so a non-perturbative term is generated in the superpotential,
$$\exp \left(-{8\pi^2}/{g_4^2}\right)\; \to \;  \exp \left( -{2\pi T}/{b_0}\right) = \exp\left(-{2\pi T}{/N}\right)=\exp\left(-aT\right)$$
where the renormalization group $\beta$-function coefficient $b_0=N$ for $SU(N)$ gauge symmetry and only holomorphic $T$ can appear in $W_\text{NP}$. Note that $N$ can be as large as hundreds or thousands from the F theory perspective \cite{Candelas:1997pq,Louis:2012nb}, and the gauge group can be semi-simple.
The KU model has a single non-perturbative term, where the $\xi$-term can lift the meta-stable ground state from AdS vacuum to a dS vacuum, but with a strong upper bound on $t=\Re{(T)}$.  This is well studied in the literature \cite{Balasubramanian:2004uy,Westphal:2006tn,Rummel:2011cd,deAlwis:2011dp,Sumitomo:2012vx} and reviewed in appendix \ref{appendix:review}.

$\bullet$ It is somewhat surprising that adding a second non-perturbative term changes the picture drastically. This is the racetrack model, also well studied in string theory. Now that we have two pure gauge groups here, namely $SU(N_1)$ for $a=2 \pi/N_1$ and $SU(N_2)$ for $b=2 \pi/N_2$ (where $N_1 \ne N_2$, otherwise they collapse to a single term). The racetrack relaxes the upper bound on $t$. On the other hand, while the stabilizing $\tau= \Im(T)$ in the single term case is trivial, stabilizing $\tau$ in the two-term racetrack case imposes strong constraint on the range of possible $\Lambda$, where, statistically, $\Lambda$ prefers to be exponentially small. More non-perturbative terms to enlarge the racetrack does not change the qualitative picture \cite{Andriolo:2019gcb}, nor does going to the Swiss-Cheese model by increasing the number of K\"ahler moduli \cite{Sumitomo:2013vla}.

$\bullet$ The remaining three terms in $V(T)$ (\ref{eq:effV}) come from the $\overline{\rm D3}$-brane and Higgs fields, respectively. The warped $\overline{\rm D3}$-brane tension explicitly breaks SUSY and gives a SUSY-breaking scale, $\tilde{T}_3 \sim {M^4_\susy}$, which are assumed to generate the various soft terms  in $S_h$ in the Higgs potential. $S_h$ tiggers the SSB in the electroweak sector and leads to an overall negative contribution to $\Lambda$. This negative Higgs contribution cancels the positive uplift from the $\overline{\rm D3}$-brane tension and allows a positive exponentially small cosmological constant. We present these three terms here:
\begin{align}
\label{V3DhSh}
	V_3&=\frac{p\tilde{T}_3}{(T + \bar{T})^{n}}\;, \quad 
	D_h =\frac{\hat{D}_h}{(T + \bar{T})^{n_D}}=\frac{\left(g^2+g'^2\right) \left({\hat v}_1^2- {\hat v}_2^2\right)^2}{8(T+\bar{T})^{n_D}},\nonumber \\
S_h&=  \frac{\hat{S}_h}{(T + \bar{T})^{n_S}}=\frac{{\hat m}_1^2 {\hat v_1}^2 + {\hat m}_2^2 {\hat v_2}^2 -2{\hat b}{\hat v}_1 {\hat v}_2}{(T+\bar{T})^{n_S}} <0\;,
\end{align}
where $V_3 \simeq m_{\susy}^4$ provides the uplift contribution from $p$ $\overline{\rm D3}$-branes. $\hat{S}_h < 0$ contains the soft terms while $\hat{D}_h > 0$ is the standard $D$-term of the Higgs potential\footnote{The tree-level model yields a Higgs mass too small to match the observed $m_h =125$ GeV. Radiative corrections as well as renormalization group flow may increase the quartic term to raise the predicted Higgs mass. In this case, we simply include them in the $D_h$ term here.}.
In string theory, all sectors are coupled (via graviton, dilaton et al) together. There is no isolated sector. Since we have assumed all modes except $T$ are stabilized already, the Higgs sectors must couple via $T$ as well. So in general, we must assume $n_D \ne 0$ and $n_S \ne 0$. It is reasonable to choose $n_D=n_S=n=2$ (see appendix \ref{appendix:barD3}), so we have
\begin{equation}
V_3 + D_h + S_h = \frac{\mathcal{D}}{(T + \bar{T})^{2}}= \frac{p\tilde{T}_3+\hat{D}_h + \hat{S}_h}{(T + \bar{T})^{2}}\;.
\end{equation}
Now, $\hat{S}_h$ contains the soft terms due to the $\overline{\rm D3}$-branes, so $\hat{S}_h \to 0$ as $\tilde{T}_3 \to 0$. In some physical sense, it is transmitted from the $\overline{\rm D3}$-brane, so it is reasonable to have $V_3 + S_h >0$ and we require $V_3 + S_h + D_h >0$. Note that we assume the presence of the soft terms in $\hat{S}_h$ without discussing how they are generated. We assume that $\mathcal{W}$ is stabilized before we determine the dynamics of $T$. Therefore, there is only one dynamical complex variable $T=t+i\tau$ in our potential. We shall remark on some variations/extension of our model below.

A preview may be useful here. Although both the $C$-term (i.e., the $\xi$-term) and the $\tilde{D}$-term (i.e., rescaled $\mathcal{D}$, see eq.\eqref{eq:model} below) will be constrained to take exponentially small values (by the factor $e^{-x}$), the small $C$-term implies an exponentially small $\mathcal{W}$ and so the value of $\Lambda$, while the small $\tilde{D}$-term implies a cancellation between the warped $\overline{\rm D3}$-brane tension and the Higgs field contributions. An exponentially small $\mathcal{W}$ yields a small $|\mathcal{W}|^{1/3} \sim {\bf m} \sim m_{\rm EW} \sim 100$ GeV, while the cancellation in $\tilde{D}$-term yields the relation $m_{\susy} \sim m_{\rm EW}$. \\
(1) If $C \propto \xi$ is turned off, the model with only an $\overline{\rm D3}$-brane and negligible $e^{-2x}$ terms has no local minimum solution, with or without the racetrack (see appendices \ref{appendix:review} and \ref{appendix:racetrack}); however, a solution exists if $n=2$ is replaced by $n=3$, but $\Lambda$ and $M_{\susy}$ are no longer as constrained. \\
(2) If the Higgs field terms $S_h$ and $D_h$ are not included, the upper bound on $\tilde{D}$ reduces to an upper bound on the $\overline{\rm D3}$-brane tension, and the SUSY-breaking scale will be negligibly small.\\
(3) If an $\overline{\rm D3}$-brane is not included, the SUSY-breaking scale from the $\xi$-term would be exponentially small; and the Higgs potential will push the ground state to an AdS vacuum with $|\Lambda| \gg \Lambda_{\rm obs}$.\\
(4) The potential $V(T)$ has the form of eq.(\ref{3termcase}) or (\ref{KUA3}) or (\ref{double}), with 3 types of terms. There are 2 branches of solutions (the $\rho_{\pm}$ similar to eq.(\ref{2sol})), in which, to a good approximation, the ``3-term" structure is reduced to a ``2-term" structure, substantially simplifying the analysis: \\
{\indent} (i) the last (uplift) term is negligible, so only AdS solutions are available;\\
{\indent} (ii) the doubly suppressed term(s) is negligible, yielding both dS and AdS solutions.\\
We are mostly interested in the branch with dS solutions.

\section{Analysis}\label{sec:analysis}

The analysis follows closely ref.\cite{Sumitomo:2013vla}.  Let us first discuss the $\alpha'^3$-correction term $\xi$ inside $V_F$.  It is easily seen from the K\"ahler potential with $\xi$ included (\ref{Modele}), that
\begin{equation}
	K^{T\bar{T}}K_T K_{\bar{T}}-3=3\xi \frac{\mathcal{V}^2+7\mathcal{V}\xi+\xi^2}{(\mathcal{V}-\xi)(2\mathcal{V}+\xi)^2}\overset{\xi\to0}{=}0\;,
\end{equation}
indicating that $\xi$ breaks the no-scale relation. The $F$-term potential could be written as \cite{Rummel:2011cd}
\begin{align}
	V_F&=\mathcal{N} e^K\left[K^{T\bar{T}}\left(\partial_T W\partial_{\bar{T}}\bar{W}+(\partial_T WK_{\bar{T}}\bar{W}+\text{c.c.})\right)+3\xi \frac{\mathcal{V}^2+7\mathcal{V}\xi+\xi^2}{(\mathcal{V}-\xi)(2\mathcal{V}+\xi)^2}|W|^2\right] \nonumber \\
	&\simeq \mathcal{N} \left[V_0(\mathcal{V},W)+\frac{\xi}{4\mathcal{V}}V_1(\mathcal{V},W)+\mathcal{O}\left(\left(\xi/\mathcal{V}\right)^2\right)\right]
	\label{eq:vf}
\end{align}
as $\xi/\mathcal{V}\ll1$ in large volume scenario. Here $V_0$ is the basic potential with a supersymmetric AdS solution and $V_1$ provides the K\"ahler uplift contribution. Explicitly, they are
\begin{align}
	V_0(\mathcal{V},W)&=\frac{1}{3\mathcal{V}^{2/3}}\partial_T W\partial_{\bar{T}}\bar{W}-\frac{1}{\mathcal{V}^{4/3}}\left(W\partial_{\bar{T}}\bar{W}+\text{c.c.}\right)\;,\nonumber\\
	V_1(\mathcal{V},W)&=\frac{1}{3\mathcal{V}^{2/3}}\partial_T W \partial_{\bar{T}}\bar{W}+\frac{1}{\mathcal{V}^{4/3}}\left(W\partial_{\bar{T}}\bar{W}+\text{c.c.}\right)+\frac{3}{\mathcal{V}^2}|W|^2\;.\label{eq:kahler}
\end{align}
Total potential $V(T)$ \eqref{eq:effV} could be expressed in the following form, in the large $t$ limit,
\begin{align}
	V(T)&\simeq \left(-\frac{a^3 A\mathcal{W}\mathcal{N}}{2}\right)\lambda(x,y)\;,\nonumber\\
	\lambda(x,y)&=-\frac{e^{-x}}{x^2}\cos{y}-\frac{\beta}{z}\frac{e^{-\beta x}}{x^2}\cos{(\beta y)}+\frac{C}{x^{9/2}}+\frac{\tilde{D}}{x^{2}}\;,\nonumber\\
	C&=-\frac{3\xi a^{3/2} \mathcal{W}}{32\sqrt{2}A}\;,\quad \tilde{D}=-\frac{\mathcal{D}}{2aA \mathcal{W}\mathcal{N}}\;,\label{eq:model}
\end{align}
where $x=at$, $y=a\tau$, $\beta=b/a$ and $z=A/B<0$ are all real parameters. We choose $A\mathcal{W} <0$, so $C >0$. Note that $\beta\neq 1$, otherwise the two non-perturbative terms collapse to one term. Without loss of generality, we let $\beta >1$.   Here $\tilde{D}$ includes the $\overline{\rm D3}$-brane as well as the two Higgs fields, as defined in eq.\eqref{V3DhSh}, and we shall consider the case $n_D=n_S=n=2$ unless stated otherwise. The different $\mathcal{W}$-dependences in $C$ and $\tilde{D}$ foretell the different roles they play in the model.

Actually the full $V(T)$ \eqref{eq:effV} includes the $\rho e^{-2x}$-like terms (i.e., the set of doubly suppressed terms) as shown in eq.(\ref{double}). As in the case of the KU model, there are two solutions to the value of $\rho$, similar to the $\rho_+$ and the $\rho_-$ given in eq.(\ref{2sol}), where $\rho_+ \gg \rho_-$ for positive (or small) $\tilde{D}$.  The $\rho_+ e^{-2x}$-like terms dominate over the uplift terms in the the $\rho_+$ solution, which yields only AdS solutions. On the other hand, the $\rho_-$ branch yields dS solutions, where the $\rho_- e^{-2x}$-like terms are small and can be neglected in $V(T)$ \eqref{eq:model}
(explained in  appendix \ref{appendix:review} and justified in appendix \ref{appendix:drop}) as well as higher order coupling terms between different contributions in the function $\lambda(x,y)$.  So we shall focus on the $\rho_-$ solution, i.e., $V(T)$ as given in  \eqref{eq:model} where the $\rho e^{-2x}$-like terms have been dropped.      

Generally, $\lambda(x,y;z,\beta,C,\tilde{D})$ is a function of $(x,y)$ with parameters $(z,\beta,C,\tilde{D})$. Our goal is to find the solution at finite $x$, if such solution exists. Solving extremal conditions $\partial_x\lambda=\partial_y\lambda=0$ yields
\begin{equation}
	y=0\;,\quad \frac{\beta}{z}=\frac{e^{\beta x}}{\beta x+2}\left[-(x+2)e^{-x}+\frac{9C}{2x^{5/2}}+2\tilde{D}\right]\;,
	\label{eq:sln}
\end{equation}
which gives solution $\lambda_\text{ext}(z,\beta,C,\tilde{D})$. There are four degrees of freedom in this system. One can rewrite all quantities as functions of $(x,\beta,\tilde{D},\lambda_\text{ext})$, where $x$ is the stabilized value. (As an alternative, we can write all quantities as functions of $(x,\beta, C, \lambda_\text{ext})$ instead. See appendix \ref{appendix:combined}). Here, the
parameters $z$ and $C$ are expressed below.
\begin{align}
	\frac{1}{z}&=\frac{e^{\beta x}}{\beta (2\beta x-5)}\left[-(2x-5)e^{-x}+9x^2\lambda_\text{ext}-5\tilde{D}\right]\nonumber\\
	C&=\frac{2x^{5/2}}{2\beta x-5}\left[(\beta-1)x e^{-x}+(\beta x+2)x^2\lambda_\text{ext}-\beta x \tilde{D}\right]\label{eq:zC}
\end{align}
We investigate the Hessian of solution which satisfy eq.\eqref{eq:sln}. Here we expand the complicated coefficients in large volume limit, $1/x\ll1$ (see appendix \ref{appendix:cal} for more detailed expressions),
\begin{align}
	\left.\partial_x^2\lambda\right|_\text{ext}&\simeq\frac{9(2\beta^2 x^2 -3\beta x-10)}{2x^2(2\beta x-5)}\Bigg[e^{-x}\frac{2(\beta-1)}{9\beta x}\left(1-\frac{5(\beta+1)}{2\beta x}+\frac{3}{2\beta x}+\cdots\right)\nonumber\\
	&\qquad\qquad\qquad\qquad\qquad\qquad+\frac{5\tilde{D}}{9x^2}\left(1-\frac{2}{\beta x}+\cdots\right)-\lambda_\text{ext}\Bigg]\;,\nonumber\\
	\left.\partial_y^2\lambda\right|_\text{ext}&=\frac{9\beta^2}{2\beta x-5}\left[-e^{-x}\frac{2(\beta-1)}{9\beta x}\left(1-\frac{5(\beta+1)}{2\beta x}\right)-\frac{5\tilde{D}}{9x^2}+\lambda_\text{ext}\right]\;,\\
	\left.\partial_x\partial_y\lambda\right|_\text{ext}&=\left.\partial_y\partial_x\lambda\right|_\text{ext}=0\;,
	\nonumber
\end{align}
where we have used eq.\eqref{eq:zC} to replace parameter $(z,C)$ with $(x,\beta,\tilde{D},\lambda_\text{ext})$. If the mass squared for both directions are positive, the solution is locally stable. This gives inequalities $m_x^2=\left.\partial_x^2\lambda\right|_\text{ext}\geq0$ and $m_y^2=\left.\partial_y^2\lambda\right|_\text{ext}\geq0$, which put strong bounds on the solution $\lambda_{\rm ext}$,
\begin{equation}
	\lambda_\text{min}\leq\lambda_\text{ext}\leq\lambda_\text{max}\;,
	\label{eq:bounds}
\end{equation}
where $m_x^2\geq0$ yields $\lambda_\text{max}$ and $m_y^2 \geq0$ yields $\lambda_\text{min}$,
\begin{align}
\label{maxmin2}
	\lambda_\text{max}&\simeq e^{-x}\frac{2(\beta-1)}{9\beta x}\left(1-\frac{5(\beta+1)}{2\beta x}+\frac{3}{2\beta x}+\cdots\right)+\frac{5\tilde{D}}{9x^2}\left(1-\frac{2}{\beta x}+\cdots\right)\;,\nonumber\\
	\lambda_\text{min}&= e^{-x}\frac{2(\beta-1)}{9\beta x}\left(1-\frac{5(\beta+1)}{2\beta x}\right)+\frac{5\tilde{D}}{9x^2} .
\end{align}
Recall that $\beta\gtrsim1$. For large $x\to \infty$, $ \lambda_\text{min} \to \lambda_\text{max}$. Define the gap as
\begin{equation}
\label{Delta}
\Delta\equiv\lambda_\text{max} - \lambda_\text{min} =  e^{-x}\frac{(\beta-1)}{3(\beta x)^2} -\frac{10\tilde{D}}{9\beta x^3}>0\;,
\end{equation}
which describes size of parameter space for allowed $\lambda_{\rm ext}$ value. Since the existence of a solution requires $\Delta>0$, this gives an upper bound for $\tilde{D}$,
\begin{equation}
	\label{eq:upD2}
	\tilde{D}\lesssim \frac{3}{10} x e^{-x}\frac{\beta-1}{\beta}\;.
\end{equation}
This is the main result of the analysis. 

We see from the above bound that order-of-magnitude-wise, a positive $\tilde{D}$ must be exponentially small to ensure a locally stable solution. In another word, if the uplift contribution is too large, including $C$ and $\tilde{D}$, there is no stable solution. This implies that the warped $\overline{\rm D3}$-brane tension must be tuned to cancel the Higgs contributions inside  $\tilde{D}$.  If $\tilde{D}<0$, solutions exist but the total potential value is not guaranteed to be positive. The positivity of minimum potential value is dependent on competition between $C$-term and $\tilde{D}$-term. Let us focus on the situation where $\lambda_{\rm min}>0$ for now. This strong upper and lower bound already ensure the fact that $\lambda_\text{ext}$ is exponentially suppressed in large $x$ limit. To estimate the potential value, we make the approximation 
\begin{equation}
	\lambda_\text{ext}\simeq\lambda_\text{min}= e^{-x}\frac{2(\beta-1)}{9\beta x}\left(1-\frac{5(\beta+1)}{2\beta x}\right)+\frac{5\tilde{D}}{9x^2}\;.
	\label{eq:lext}
\end{equation}
We keep the most important part for $e^{-x}$-term and $\tilde{D}$-term, and neglect higher order term in $1/x$. Careful treatment in appendix \ref{appendix:cal} shows that at leading-order, this approximation is solid. With eq.\eqref{eq:lext}, we could reduce one degree of freedom in all quantities and express parameters as functions of $(x,\beta,\tilde{D})$. Now we have three degrees of freedom in the system. Eq.\eqref{eq:zC} become
\begin{equation}
	\frac{1}{z}\simeq - \frac{e^{-(1-\beta)x}}{\beta^3}\;,\quad	C\simeq x^{7/2}e^{-x}\frac{2(\beta-1)}{9\beta}-\frac{4}{9}x^{5/2}\tilde{D}\;.
	\label{eq:cz}
\end{equation}
With this choice of $\lambda_{\rm ext}$ approximation, $z$ depends on $x$ and $\beta$ only, which is convenient for statistical analysis. $C$ is related to $\mathcal{W}$ through $\xi$ as shown in eq.\eqref{eq:model}, from which we obtain
\begin{equation}
\label{Wvalue}
	\mathcal{W}\simeq-\frac{32 A}{27\xi} \left(\frac{2x}{a}\right)^{3/2}  \left[x^2 e^{-x} \frac{(\beta -1)}{\beta} -2x\tilde{D}\right]
\end{equation}
By considering the upper bound on $\tilde{D}$, we see that $\mathcal{W}\sim e^{-x}$ as mentioned in the introduction. Recall that $\mathcal{W}$ in the model is {\it a priori} independent of $x$. In scanning over the discrete flux parameters in the landscape,  $\mathcal{W}$ takes a wide range of values. A minimum solution exists only for $\mathcal{W}$ with values given by eq.(\ref{Wvalue}).  Therefore, cosmological constant $\Lambda$, which is the minimum value of potential, could be expressed as a function of $(x,\beta,\tilde{D})$,
\begin{align}
	\Lambda&=V_\text{min}\simeq\left(-\frac{a^3 A \mathcal{W}\mathcal{N}}{2}\right)\lambda_\text{ext}\nonumber\\
	&\simeq \frac{256A^2\mathcal{N}}{243\xi}\left(\frac{a}{2x}\right)^{3/2}\left[x^2e^{-x}\frac{\beta-1}{\beta}-2x\tilde{D}\right]\left[x^2e^{-x}\frac{\beta-1}{\beta}+5x\tilde{D}\right]\;,
	\label{eq:cc}
\end{align}
So $\lambda$ and $\mathcal{W}$ go like $ e^{-x}$ and $\Lambda$ goes like $e^{-2x}$. Rearrange a little and we arrive at the expression
\begin{equation}
	\Lambda\approx \frac{3\mathcal{N}\xi \mathcal{W}^2}{4(2t)^{9/2}}+\frac{\mathcal{D}}{(2t)^{2}}\;.
	\label{eq:cc2}
\end{equation}
Because of the upper bound for $\tilde{D}$ in eq.\eqref{eq:upD2}, there exists an upper bound for $\mathcal{D}$, which is
\begin{equation}
\label{upperD2}
	\frac{\mathcal{D}}{(2t)^2}=\frac{\tilde{T}_3+\hat{D}_h+\hat{S}_h}{(2t)^2}\lesssim \frac{81\mathcal{N}\xi\mathcal{W}^2}{32(2t)^{9/2}}\;.
\end{equation}
As we shall discuss below, in the absence of fine-tuning, both K\"ahler modes have exponentially small masses \cite{Tye:2016jzi}.

If we let there be $M$ such uplift contributions (other than K\"ahler uplift) in the potential with different coupling power $n_i$, the bound (\ref{eq:upD2}) becomes 
\begin{equation}
	 \sum_i^M\frac{n_i \tilde{D}_i}{x^{n_i}}\left(1-\frac{2n_i}{9}\right)\lesssim e^{-x} \frac{\beta-1}{3\beta x}\;,
	\label{eq:upDi}
\end{equation}
which is considered in appendix \ref{appendix:cal}. If $n_D=n_S=n=3$, which is also possible as we discussed in last section, eq.\eqref{eq:upDi} simplifies to 
\[
\tilde{D}\lesssim x^2 e^{-x} \frac{\beta-1}{3\beta}\;.
\]
The cosmological constant in the general case is presented in eq.\eqref{eq:generalL} and eq.\eqref{eq:generalLL}. General upperbound for $\sum \mathcal{D}_i/(2t)^{n_i}$ is shown in eq.\eqref{eq:upd} (under some conditions). For $n_D=n_S=n=3$, this upper bound (\ref{upperD2}) for $\mathcal{D}/(2t)^3$ becomes
\begin{equation}
	\frac{\mathcal{D}}{(2t)^3}=\frac{\tilde{T}_3+\hat{D}_h+\hat{S}_h}{(2t)^3}\lesssim \frac{9\mathcal{N}\xi\mathcal{W}^2}{8(2t)^{9/2}}\frac{\beta x}{2(\beta+1)}\;.
\end{equation}
There is an extra $x$ factor in the upper bound, indicating that different $n_i$ would give slightly different result and order-of-magnitude wise, this $\sum \mathcal{D}_i/(2t)^{n_i}$ factor is limited by the $\xi$-term. This tells us that to ensure the existence of solutions, the additional contribution, apart from RKU should not be large. The observed cosmological constant is exponentially small, which implies that this sum essentially  vanishes. This is possible given the fact that $(D_h+S_h)$ is negative, which could cancel the positive contribution $V_3$. If we have $\overline{\rm D3}$-brane alone (without K\"ahler uplift), there is no reason why $\Lambda$ is naturally small. It is the combination of RKU and $\overline{\rm D3}$-brane together with the Higgs doublets that can yield a solution with a naturally exponentially small $\Lambda$ and a SUSY-breaking scale comparable to $m_{\rm EW}$.

\section{Statistical Preference for a Small Positive $\Lambda$}\label{sec:sta}

Although the exponential factor in eq.\eqref{eq:bounds} suggests that $\lambda$ (and so $\Lambda$) can be exponentially small, {\it A priori}, it does not have to. To see quantitatively that an exponentially small positive $\Lambda$ is statistically preferred, let us see what happens if we put random values for the parameters in the model and find the probability distribution $P(\Lambda)$ of $\Lambda$, in particular for $\Lambda \sim 0^+$. We shall justify the large $x$ approximation {\it A posteriori} as this simplifies the discussion.  The analysis follows ref.\cite{Sumitomo:2013vla} and details can be found in appendix \ref{appendix:stat}. Let us start with eq.\eqref{eq:cc}, \eqref{eq:cc2} and \eqref{upperD2}. The order of magnitude of $\Lambda$ is dictated by $x$, the stablized value at the minimum potential. However, $x$ is an output parameter. So we should trade it for the input parameter $z$ and other parameters in the model. It turns out $P(\Lambda)$ is sensitive to $z$ and most sensitive to $\beta$. So we shall focus on these two parameters.

Recall that $\tilde{D}$ is essentially the sum of terms in $V(T)$ \eqref{eq:effV} outside the $V_F$.
Let us introduce a factor $q$ so the bound (\ref{eq:upD2}) is expressed as an equality,
\begin{equation}
\label{qD}
	\tilde{D}= q \frac{3}{10} x e^{-x}\frac{\beta-1}{\beta}, \quad \quad q<1\;.
\end{equation}

\subsection{Positive $\tilde{D}$}

Now, let us first assume $\tilde{D} >0$, so $q>0$ and $\Lambda>0$. We construct $\hat{\Lambda}$ by referring to $\Lambda(x,\beta,\tilde{D})$ (\ref{eq:cc}),
\begin{align}
	\label{eq:hL}
	\hat{\Lambda} &\simeq \frac{256A^2\mathcal{N}}{243\xi}\left(\frac{a}{2x}\right)^{3/2}x^2\left[xe^{-x}\frac{\beta-1}{\beta}-2\tilde{D}\right]\left[xe^{-x}\frac{\beta-1}{\beta}+5\tilde{D}\right]\; \nonumber\\
& \simeq \frac{2^{13/2}}{3^5}\frac{(\beta-1)^2}{\beta^2}x^{5/2}e^{-2x}\left(1-\frac{3}{5}q\right)\left(1+\frac{3}{4}q\right)\nonumber\\
	&\simeq \frac{2^{13/2}}{3^5}\frac{\kappa^{\frac{2}{\beta-1}}}{\beta^2\sqrt{\beta-1}}\left(-\ln{\kappa}\right)^{5/2}\left(1-\frac{3}{5}q\right)\left(1+\frac{3}{4}q\right)\nonumber\\
	&\overset{q\to0}{=}\frac{2^{13/2}}{3^5}\frac{\kappa^{\frac{2}{\beta-1}}}{\beta^2\sqrt{\beta-1}}\left(-\ln{\kappa}\right)^{5/2}\;,
\end{align}
where, in the second line we suppress the factors $|A|, a, \mathcal{N}$ and $\xi$ to focus on the dependence on $(x,\beta)$ and apply eq.(\ref{qD}) to replace $\tilde{D}$ satisfying eq.\eqref{eq:upD2}. We shall keep $q$ as a real parameter here. In the third line, we replace $x$ by $z=A/B<0$, or by $\kappa$, using eq.\eqref{eq:cz}, where
$$ \kappa \equiv \frac{- z}{\beta^3} = e^{-(\beta -1)x}\;.$$
so $\beta^3 > \kappa >0$. The dependence of $q$ affects little on $\hat{\Lambda}$. This is the reason we simply let $q=0$ in the last line.

Now we investigate the probability distribution $P(\Lambda)$. There are complicated dependence of $\beta$ in $\hat{\Lambda}$, so let us first find $P(\Lambda; \beta)$ as $\Lambda \to 0^{+}$,
\begin{align}
	P(\Lambda; \beta) &= \int {\rm d} z P(z) \delta\left(\Lambda - {\hat \Lambda}(z)\right)\nonumber\\
	&\overset{\Lambda\sim 0^{+}}{\simeq}2^{1-2\beta}\cdot 3^{\frac{5(\beta-1)}{2}}\cdot \frac{\beta^{\beta+2}}{(\beta-1)^{\beta-2}}\frac{1}{\Lambda^{\frac{3-\beta}{2}}(-\ln{\Lambda})^{\frac{5(\beta-1)}{4}}}\;,
\end{align}
in which ${\hat \Lambda}(z)$ is given in eq.\eqref{eq:hL} and we take $P(z)$ to be flat, $P(z)=1$ for $z$ in the parameter space $-1 \le z \le 0$.

Next we consider $\beta=b/a = N_1/N_2 >1$, where $a=2\pi/N_1$ for gauge group $SU(N_1)$ and $b=2\pi/N_2$ for gauge group $SU(N_2)$, both from D7-branes. Scanning through $N_1=3, 4,\dots,N_{\rm max}$ and $N_2=2, 3,\dots, (N_1 -1)$.  The most divergent $P(\Lambda;\beta)$ at $\Lambda=0^{+}$ is when $(3-\beta)/2 = 1 -(\beta -1)/2$ is closest to unity, namely, the smallest $\beta >1$: $\beta_{\rm min} =N_{\rm max}/(N_{\rm max}-1)$, where $N_{\rm max} \gg 2$.
So, to a good approximation, we may set $P(\Lambda)$ to the most divergent $P(\Lambda;\beta)$, i.e.,
\begin{equation}
\label{Papprox}
P(\Lambda)\simeq P(\Lambda;\beta_{\rm min})\overset{\Lambda \sim 0^+}{\sim} k\Lambda^{-1+k}, \quad k=\frac{1}{2(N_{\rm max}-1)}\;.
\end{equation}
Here we have neglected the $(-\ln{\Lambda})$-factor for large $N_{\rm max}$. To summarize, the divergent behavior of $P(\Lambda)$ comes from scanning over $z$ while the strength of the divergence comes from scanning over $\beta$. They are precisely the two parameters that measure the relative roles of the two non-perturbative terms in the racetrack. 

Here $P(\Lambda)$ around $\Lambda \sim 0^{+}$ is normalizable for any finite $N_\text{max}$, but not at $\Lambda \to \infty$. In any case, our model breaks down for $\Lambda > M_{\rm Pl}^4$. So we need to introduce a cut-off. There are a number of ways to do so. For convenience sake, we may simply set $P(\Lambda)=0$ for $\Lambda >1$, 
\[
P(\Lambda)=
\begin{cases}
k \Lambda^{-1+k}\;, & 0<\Lambda\leq 1\\
0\;,&\Lambda>1
\end{cases}
\;,\quad \quad \int_0^{1}P(\Lambda){\rm d}\Lambda =1\;.
\]
It is useful to introduce $\Lambda_{Y\%}$, which means that $Y \%$ of the probability that has a $\Lambda < \Lambda_{Y\%}$. So $\Lambda_{50\%}$ is the median. One finds that $N_\text{max} \simeq 201$ if we match $\Lambda_{50 \%}$ to the observed $\Lambda_\text{obs}$.  One gets  $N_{\rm max} \simeq 62$ if we let $\Lambda_{10 \%} = \Lambda_\text{obs}$ \cite{Andriolo:2018dee}. These are order-of-magnitude estimates. Following F theory perspective, Type IIB string theory can certainly accommodate such a value for $N_\text{max}$ \cite{Candelas:1997pq,Louis:2012nb}. 

It is clear that the determination of $N_{\rm max}$ is somewhat sensitive to the details of the cut-off; 
 the value of $N_{\rm max}$ will change if we adopt a different cut-off on $\Lambda$. 
As an example, if we use the Weibull distribution as an alternative way for cut-off (see  appendix \ref{appendix:stat}). In this case, matching $\Lambda_{50 \%}$ to $\Lambda_\text{obs}$ yields $N_{\rm max} \sim 380$. $N_{\rm max}$ will also be different if we add additional non-perturbative terms to the racetrack \cite{Andriolo:2019gcb}.  Fortunately, the qualitative features do not change, and the determination of ${\bf m}$ or $m_{\rm EW}$ is a lot less sensitive. The overall picture is robust.

\subsection{Negative $\tilde{D}$}

What happens if $\tilde{D}<0$ ? This means that the negative soft term $S_h$ overcomes the positive terms $V_3$ and $D_h$, so $V_3+S_h+D_h <0$. This is unlikely if $S_h$ is transmitted from the warped $\overline{\rm D3}$-brane tension. However, this is entirely possible if the $S_h$ term comes from a random sampling of the fluxes, as we believe is in the analogous case for the quark and lepton masses \cite{Andriolo:2019gcb}. Recall eq.\eqref{maxmin2} and eq.\eqref{Delta} for $\tilde{D}<0$,
\begin{align}
\lambda_\text{min}&\simeq e^{-x}\frac{2(\beta-1)}{9\beta x}-\frac{5|\tilde{D}|}{9x^2}\;,\nonumber \\
\Delta&= \lambda_\text{max} - \lambda_\text{min} \simeq  e^{-x}\frac{(\beta-1)}{3(\beta x)^2} +\frac{10|\tilde{D}|}{9\beta x^3}\;.
\label{minusD}
\end{align}
If $-4/3<q<0$, in which $|\tilde{D}|$ remains negligibly small so $\Lambda$ is still positive and goes like $e^{-2x}$, statistical property for $\Lambda$ stays the same as the case for positive $\tilde{D}$. Note that the  $|\tilde{D}|$ term in eq.(\ref{minusD}) does not necessarily have the $e^{-x}$ factor. For large $x$ and/or large $|\tilde{D}|$, the $\tilde{D}$-term dominates over the $e^{-x}$-term and so $\lambda_{\rm ext} <0$ and the allowed range for $\lambda_{\rm ext}$ (i.e., $\Delta$) also grows relative to the exponentially suppressed term. As we vary the parameters (i.e., scanning over the various parameters), $P(\Lambda)$ is smooth even as $\Lambda \to 0^-$.  This is similar to the case for the KU Model or the $\overline{\rm D3}$-brane model, where $P(\Lambda)$ is smooth as $\Lambda \to 0$, except here, this is the case only as $\Lambda \to 0^-$, not when $\Lambda \to 0^+$. The resulting $P(\Lambda)$ is sketched in figure \ref{fig:proL}. Similar behavior is obtained in the $\rho_+$ branch of AdS solutions. So we conclude that an exponentially small positive (but not negative) $\Lambda$ is preferred. 

In the absence of the $\overline{\rm D3}$-brane, $S_h+D_h \simeq - m^4_{\susy} <0$, which will yield an unacceptably large negative $\Lambda$, as well as the undesirable spectrum of particle-super-partner mass denegeracy.

\section{Supersymmetric Standard Model}\label{sec:phe}

The model (\ref{eq:effV}) has a number of parameters where the potential $V(T)$ \eqref{eq:model} is totally determined by the input parameters $(A, z=A/B, a,\beta=b/a,C, \xi,\mathcal{N})$ (where $C\propto \mathcal{W}/A$ \eqref{eq:model}), in addition to $\tilde{D}$ in eq.\eqref{eq:model}. In the statistical analysis, equating the median $\Lambda_{50\%}$ to the observed  $\Lambda_{\rm obs}$ \eqref{Lambda} determines $N_{\rm max} \simeq 201$, which determines $a=2\pi/N_{\rm max}$ and $\beta=N_{\rm max}/(N_{\rm max}-1)$. Parameters $\xi$ and $\mathcal{N}$ are determined before the stablization and they are prefactors in $\Lambda$. We know that $|A|\sim\mathcal{O}(1)$. If we choose a value for $A$, express $\tilde{D}$ using eq.\eqref{qD} (fix the $q$) and describe $\Lambda_{\rm obs}$ by eq.\eqref{eq:cc}, we arrive at a single equation where $x$ can be solved. Then one can use eq.\eqref{eq:zC} to solve for parameter $z$ and $C$. A typical choice
$$\xi \simeq 10^{-2}, \quad \mathcal{N} \simeq 10^{-3}, \quad A=-1,\quad q=0, \quad N_{\rm max} =201, \quad \beta=1.005,\quad a=0.03126$$  yields, 
$$x=134.7, \quad z=-0.5175,  \quad C=9.905 \times 10^{-55} , \quad  \lambda_{\rm ext} = 2.422\times 10^{-64} $$
so $$2t=8.620\times 10^{3} , \quad M_S=2.683 \times 10^{15}\;{\rm GeV}, \quad \mathcal{W} =2.7 \times 10^{-49}.$$
These results hardly change as $\lambda_\text{ext}$ varies between $\lambda_\text{min}$ and $\lambda_\text{max}$, simply because $\lambda_\text{min}$ is very close to $\lambda_\text{max}$. Overall, $\Lambda$ is most sensitive to the value of $N_{\rm max}$, followed by the value of $z$.
Different choices of $A$ , $\xi$, $\mathcal{N}$ and $q$ can lead to different $x$ (e.g., changing $A$ by a factor of 10, or $\mathcal{N}$ by a factor of 100, shifts $x$ by less than $2\%$). Only narrow ranges of $z$ and $C$ (representing the flux parameters in the landscape) can yield meta-stable dS solutions. 

\subsection{The Electroweak Scale}

Now let us rewrite eq.\eqref{eq:cc2} in terms of the emerging mass scale ${\bf m}$,
\begin{equation}
{\bf m} = {|\mathcal W|}^{1/3} \simeq (2t)^{3/4} \left(\frac{4 \Lambda_{\rm obs}}{3 \xi \mathcal N}\frac{1-\frac{3}{5}q}{1+\frac{3}{4}q}\right)^{1/6}\;.
\end{equation}
Using above numerical value for parameters and assuming $|W_h|=|\mu h_1h_2|$ is  of the same order of magnitude as ${\mathcal W}=W_h + W_0(U_i, S)$, one could get
\begin{equation}
	m_\text{EW}\simeq {\bf m} \sim|\mathcal{W}|^{1/3}\;M_\text{Pl}\sim 155.2 \;\text{GeV}\;,
\end{equation}
If $|W_h|$ is small compared to $W_0(U_i, S)$, then $m_{\rm EW} < {\bf m}$ in the absence of fine-tuning. In any case, this may be considered as a resolution to the $\mu$ problem, i.e., why $\mu \sim m_{EW}$ and not of order of the Planck scale. 

Because of the statistical nature of the analysis, it is hard to determine $m_{\rm EW}$ more precisely. For example, if we choose $\Lambda_{10\%}$ to match the  $\Lambda_{\rm obs}$, $N_{\rm max} \simeq 64$ and $2t \simeq 2.9 \times 10^{3}$, which yields $m_{\rm EW} \simeq 43$ GeV. Enlarging the racetrack to 3 or more non-perturbative terms (instead of 2 terms) tends to decrease $m_{\rm EW}$, while adopting a smaller ${\mathcal N}$ increases $m_{\rm EW}$. So let us take, up to a factor of 2 or 3,
\begin{equation}
m_{\rm EW} \sim 100 \, {\rm GeV}\;,
\end{equation}
$|W_h|>0$ indicates that Higgs fields acquired vev's. Here, the vev's $v_i \ne 0$ implies that SSB has taken place. To have vev's of order of $m_{\rm EW}$ implies that the other terms in the Higgs potential are of similar orders of magnitude. Let us discuss this in some detail.

\subsection{Relation between EW scale and SUSY-breaking scale}

In SSM phenomenology, one considers two $SU(2)$ Higgs-doublets $h_1=(h_1^{+},h_1^0)$ and $h_2=(h_2^0,h_2^{-})$, with the Higgs potential,
\begin{align}
	V_h&=\left(|\mu|^2+m_1^2\right)h_1^\dagger h_1+\left(|\mu|^2+m_2^2\right)h_2^\dagger h_2 +\left(b h_1h_2+\text{c.c.}\right)\nonumber\\[3pt]
	&\qquad+\frac{g^2}{8}\left(h_1^\dagger\vec{\sigma}h_1+h_2^\dagger\vec{\sigma}h_2\right)^2+\frac{g'^2}{8}\left(h_1^\dagger h_1-h_2^\dagger h_2\right)^2\;,
\end{align}
where $bh_1h_2\equiv b\left(h_1^{+} h_2^{-}-h_1^0 h_2^0\right)$ and $\vec{\sigma}$ is the Pauli matrices. Because of the coupling between Higgs field and K\"ahler modulus, the Higgs fields ${\hat h}_i$ from the string theory perspective has to be rescaled to $h_i$ that appears in SSM phenomenology, $h_i=X\hat{h}_i$. Instead of $n_D=n_S=n=2$, let us consider the case with slightly more general $n_D$ and $n_S$. 

Let us discuss the $D$-term potential first,
\begin{equation}
	\hat{D}_h=\frac{g^2}{8}\left(\hat{h}_1^\dagger\vec{\sigma}\hat{h}_1+\hat{h}_2^\dagger\vec{\sigma}\hat{h}_2\right)^2+\frac{g'^2}{8}\left(\hat{h}_1^\dagger \hat{h}_1-\hat{h}_2^\dagger \hat{h}_2\right)^2\;,
\end{equation}
After K\"ahler modulus $T$ stablized, $D$-term potential we observed is 
\begin{equation}
	\frac{\hat{D}_h}{(2t)^{n_D}}=\frac{1}{X^4(2t)^{n_D}}\left[\frac{g^2}{8}\left(h_1^\dagger\vec{\sigma}h_1+h_2^\dagger\vec{\sigma}h_2\right)^2+\frac{g'^2}{8}\left(h_1^\dagger h_1- h_2^\dagger h_2\right)^2\right]\equiv D_h\;.
\end{equation}
Therefore, the rescale factor is determined as $X=1/{(2t)^{n_D/4}}$. Note that we choose not to rescale the gauge couplings, since their values are close to the string coupling $g_s =1/{\Re(S)}$ (see  appendix \ref{appendix:prefactor}). If we choose to rescale them, the power of $(2t)$ involved will have to be rather small, so it has little impact on our overall picture.

Higgs fields in the superpotential are also rescaled as
\begin{equation}
	\hat{W}_h=\hat{\mu}\hat{h}_1\hat{h}_2=\frac{\hat{\mu}}{X^2} h_1 h_2=\mu h_1 h_2\equiv W_h\;,
\end{equation}
where $\mu$ is the rescaled value that appears in SSM. Soft terms for SSM Higgs after rescaling is
\begin{equation}
	\frac{\hat{S}_h}{(2t)^{n_S}}=\frac{1}{(2t)^{n_S-n_D/2}}\left[\hat{m}_1^2 h_1^\dagger h_1+\hat{m}_2^2 h_2^\dagger h_2+\left(\hat{b} h_1 h_2+\text{c.c.}\right)\right]\equiv S_h\;.
\end{equation}
Thus, parameters in SSM Higgs are rescaled as
\begin{equation}
	\mu=\hat{\mu}(2t)^{0.5n_D}\;,\quad m_i^2=\hat{m}_i^2(2t)^{0.5n_D-n_S}\;,\quad b=\hat{b}(2t)^{0.5n_D-n_S}\;.
\end{equation}
We are interested in classical solution where $|h_1^{+}|=|h_2^{-}|=0$ and $|h_1^{0}|=v_1$, $|h_2^{0}|=v_2$, and we choose $b$ to be real and positive. The potential is bounded from below along any direction, including $v_1=v_2$, if $2|\mu|^2+m_1^2+m_2^2>2{b}$. SSB will take place if there is a tachyonic direction at the origin, or  
\begin{equation}
	\det{\left.V_h''\right|_{v_i=0}}<0 \quad \to \quad \left(|\mu|^2+m_1^2\right)\left(|\mu|^2+m_2^2\right)<{b}^2\;.
\end{equation}
After SSB, the potential at the local minimum is negative, $\left.V_h\right|_{\rm min}<0$, since $V_h|_{v_i=0}=0$ as shown in eq.\eqref{Higgs1}. As other terms are positive, soft terms $S_h$ must be negative,
\begin{equation}
S_h= m_1^2v_1^2 + m_2^2v_2^2 -2bv_1v_2 <0\;.
\end{equation}
In the $\mathcal{D}\geq0$ case, it is bounded by the value of $\Lambda_{\rm obs}$, which is exponentially small;  in phenomenongy, it is an excellent approximation to take  
\begin{equation}
V_3+ S_h +D_h \simeq 0\;.
\label{eq:sum}
\end{equation}
$\tilde{T}_3$ in $V_3$ is the warped tension of the $\overline{\rm D3}$-brane, which is responsible of SUSY breaking. So it is reasonable to adopt
\begin{equation}
\tilde{T}_3=M_{\susy}^4\;.
\end{equation}
With the summation relation eq.\eqref{eq:sum}, we find that
\begin{equation}
	M_\susy^4\simeq\frac{(2t)^n}{n}\left(1-\frac{2n}{9}\right)^{-1}\left[n_S\left(1-\frac{2n_S}{9}\right)\left|S_h\right|-n_D\left(1-\frac{2n_D}{9}\right)D_h\right]\;,
\end{equation}
In the special case where $n_D=n_S=n$, we have ($v^2=v_1^2 +v_2^2$)
\begin{equation}
M_\susy^4 = (2t)^{n}\left(|S_h| -D_h\right) = (2t)^{n}\left(|\mu|^2v^2 -\left.V_h\right|_{\rm min}\right) \simeq (2t)^{n} m_{\rm EW}^4\;.
\end{equation}
With $2t \simeq 10^4$, this means 
\begin{equation}
M_\susy \sim 100 \, m_{\rm EW}\;,
\end{equation}
if $n=2$ or $M_\susy \sim 10^3 \, m_{\rm EW}$ if $n=3$.
The value changes only a little if $n_D \ne n$ and/or $n_S \ne n$ as long as $9/2 > n_D >0$ and $9/2 >n_S>0$.

In terms of the effective potential, it is $m_\susy^4 = \tilde{T}_3/(2t)^{n}$ that comes in to be canceled by the Higgs terms, so
\begin{equation}
m_\susy \simeq m_{\rm EW}\;.
\end{equation}
One may view $M_\susy$ to be the SUSY-breaking scale in the landscape while $m_\susy$ as the SUSY-breaking scale in the visible sector responsible for a dS solution in the model.  

Overall, in this model including two kinds of uplift contributions, exponentially small cosmological constant requires EW scale to be $100\;\text{GeV}$ and SUSY-breaking scale $m_\susy$ in the visible sector equals to the EW scale. The cancellation between contribution from a $\overline{\rm D3}$-brane and that from Higgs sector is the condition to have a locally stable solution in the presence of $\alpha'^3$-correction. With such a low $m_\susy$, this model predicts that the masses of the superpartners should be within the experimental reach in laboratories.

\subsection{Comments}

A few comments are in order here : 

$\bullet$ We have assumed that only soft terms are present. For the Higgs potential, this is reasonable, as additional terms are suppressed by powers of $M_{\rm Pl}$ scale. However, are non-soft interaction terms present in our model ?
At first sight, since the SUSY-breaking $\overline{\rm D3}$-brane term enters explicitly, non-soft terms may be unavoidable. However, if the KKLT uplift to dS comes from spontaneous SUSY breaking, then only soft SUSY breaking terms are expected. This may be realized via introducing flux in $D7$-branes thus generating a Fayet-Iliopoulos $D$ term \cite{Burgess:2003ic}; as an alternative, one may treat the presence of the $p$ $\overline{\rm D3}$-branes as an excited state of an AdS supersymmetric vacuum \cite{Kachru:2002gs}, in which case it can be realized as a $F$-term spontaneous SUSY breaking in the non-linearly realized Volkov-Akulov SUSY framework \cite{Ferrara:2014kva,Kallosh:2014wsa}; following this direction, one may reasonably believe that only soft SUSY-breaking terms appear. 

$\bullet$ In ref.\cite{Tye:2016jzi}, it is argued that the probability distribution $P(\Lambda)$ for $\Lambda$ is insensitive to radiative corrections, that is, it is radiative stable. The same argument should apply to the probability distribution of $P({\bf m})$ for ${\bf m}$, or equivalently to $m_{\rm EW}$, though radiative corrections and renormalization group flow can be important to the particle spectrum and couplings. This implies that, in actual practice for the purpose of phenomenology, only soft terms should be introduced in our model. Clearly it is important to understand this issue better.

$\bullet$ Will chiral symmetry breaking in QCD shift the vacuum energy density ? In this case, we should replace $V_h$ (\ref{Higgs1}):
$V_h \to V_{\rm SSM} =V_h +V_{\rm QCD}$, 
which will shift $m_{\susy}$ by a tiny amount. On the other hand, $M_{\susy}$ can be raised by  orders of magnitude if there is another scalar mode $\varphi$ with potential $V(\varphi)$ that undergoes SSB at a high energy scale; in this case, the condition $\mathcal{D} \simeq 0$ implies $\tilde{T}_3$ can be much larger. Such a contribution must depend on $(T+ \bar{T})$ to preserve the naturally small feature of $\Lambda$.

$\bullet$ The region (the basin of attraction) in the potential $V(x,y)$ that the universe will roll towards the local minimum is very small. Although it can roll towards $x \to \infty$ (i.e., decompactified to ten-dimensional spacetime) for $\lambda (x,y) > \lambda_{\rm max}$, it is more likely, for $y \ne 0$, to roll towards small $x$, and reach a value where the approximate effective potential $V(T)$ (\ref{eq:model}) is no longer valid  \cite{Andriolo:2019gcb}.

\section{Discussion and Remarks}\label{sec:dis}

Let us address some points of the framework here:

$\bullet$ In the absence of fine-tuning, the mass of K\"ahler mode $t$ is constrained by
\[
\frac{m_t^2}{\Lambda_{\rm obs}} \simeq \frac{\partial_t^2 V}{2K_{T\bar{T}}\Lambda_{\rm obs}}=\frac{2x^2\left.\partial_x^2\lambda\right|_{\rm ext}}{3\lambda_{\rm ext}}\leq\frac{9}{2}\cdot\left(1+\frac{35(\beta+1)}{6\beta x}+\cdots\right)\;.
\]
Thus, order-of-magnitudely one could estimate that 
$$ m_t \sim \sqrt{\frac{\Lambda_{\rm obs}}{M_{\rm Pl}^2}} \sim 10^{-33} {\rm eV}\;,$$
and similarly for the axionic mode $\tau$ \cite{Tye:2016jzi}. As the universe expands, when $H > m_{\tau}$, the mass is negligible compared to $H$ and the energy density stored in the potential plays the role of a dark energy, contributing to $\Lambda$. As the universe expands, $H$ decreases. When $H <m_{\tau}$, the field starts oscillating around the bottom of the potential due to the mis-alignment, converting the potential energy to dark matter density. However, for such a light mass, it is possible this mis-alignment mechanism has not yet begun (or still at an initial stage of the oscillation period, with the period comparable to or longer than the age of our universe) \{cf.\cite{Kolb:1990vq,Marsh:2015xka}\}. On the other hand, we do not rule out the possibility that another boson, be it a dilaton-axion or a complex structure modulus mode, with mass $m \sim 10^{-22}$ eV, plays the role of fuzzy dark matter that is important for structure formation in our universe.  

$\bullet$ In SSM phenomenology, the usual approach starts with SUSY breaking, which is then transmitted to the visible sector. The soft terms are generated, which leads to SSB of the Higgs sector. In the model here, the back reaction is crucial, so the coupling of the Higgs sector to SUSY breaking must be treated with SUSY breaking simultaneously. On the other hand, how exactly the soft terms are generated does enter in our simple model.  For all practical purposes, they can appear simply because of the flux choices.

$\bullet$ In the brane world scenario, our standard model particles are taken to be open string modes living in the D3/D7 branes in a warped throat while the $\overline{\rm D3}$-brane lives in another warped throat, where it is meta-stable. We note that the explicit SUSY-breaking scale is very close to the EW scale,  $m_{\susy} \sim m_\text{EW}$. This implies very similar warp factors for the two distinct throats, which requires a fine-tuning. Naturalness suggests that they actually live in the same throat. If this is the case, this also enables a stronger transmission of SUSY breaking from the $\overline{\rm D3}$-brane to the Higgs sector. This can happen in two ways: \\
(1) the stack of $\overline{\rm D3}$-branes that uplift the vacuum to a dS space also contain the standard model particles as open string modes in the $\overline{\rm D3}/D7$ system \cite{GarciadelMoral:2017vnz,Cribiori:2019hod,Parameswaran:2020ukp}; that is, the same stack of  $\overline{\rm D3}$-branes help to uplift the vacuum to a dS vacuum and uplift the SUSY scale to a reasonable value as well as provide the particle spectrum and couplings in the SSM; this attractive scenario deserves further study;   \\
(2) we have both $D3$-branes and $\overline{\rm D3}$-branes in the same throat.
Consider a throat such as the Klebanov-Strassler throat \cite{Klebanov:2000hb}. It is a deformed $R \times S^2 \times S^3$ conifold, where the $S^3$ has a size while $S^2$ stays as a point at the bottom of the throat. It is known that SUSY is broken in a resolved conifold where the $S^2$ at the bottom has a finite size. Since a 
$\overline{\rm D3}$-brane breaks SUSY explicitly, a deformed and resolved conifold is likely to be the case here. In such a throat, with an attractive force between the $\overline{\rm D3}$-brane and the stack of D3/D7 branes, a barrier is necessary to prevent them from colliding and annihilating the $\overline{\rm D3}$-brane.  

To see how viable this picture can be, let us assume there is no barrier so that the $\overline{\rm D3}$-brane yields a scenario much like the brane inflation in early universe \cite{Dvali:1998pa,Dvali:2001fw,Burgess:2001fx}: D3-$\overline{\rm D3}$ brane inflationary scenario \cite{Kachru:2003sx} or a D3/D7 scenario \cite{Koyama:2003yc,Dasgupta:2004dw}, except that the energy scale is much lower now. With warped brane tension of order of $m^4_{\susy}$,
and the Hubble value $H \sim 10^{-60} M_{\rm Pl}$,  one e-fold of accelerated expansion takes about $H^{-1}$, which is about the age of our universe. Since the number of e-folds can be $N_e > 1$, this scenario may work even in the absence of any barrier, as long as the $S^3 \times S^2$ bottom of the throat is big enough.  Clearly, the scenario would be more robust if the throat bottom is somewhat non-trivial. It will be interesting to examine the structure of the bottom of a deformed/resolved conifold.

$\bullet$ As noted earlier, the allowed range of $\Lambda$, $\Lambda_\text{min} < \Lambda < \Lambda_\text{max}$, is very small. At first sight, this seems to imply that a dS vacuum solution is highly unlikely. Does this mean we have to invoke some form of the Anthropic Principle to justify the choice? However, this is a measure issue (that is, the choice of the probability distribution of the flux parameters and the variables/parameters) which we have nothing to add.

On the other hand, if the early universe starts with multiple D3-branes and $\overline{\rm D3}$-branes, such that inflation is driven by some  D3-brane/$\overline{\rm D3}$-brane interactions. During inflation, the size of the universe, i.e., the volume (i.e., the three spatial dimensions) of the D3-branes grows exponentially. Inflation ends after the annihilation of some pairs of D3-branes and $\overline{\rm D3}$-branes. The pairs with large attractive force will collide early, while the pairs with weak attractive force (i.e., low tension) will collide late. So it is reasonable if a $\overline{\rm D3}$-brane (or a few) are left over to today, if the attractive force between the $\overline{\rm D3}$-brane and the D3/D7-branes is too weak to bring them to collide within the age of the universe, or to overcome some barrier that is present at the bottom of the throat. 

$\bullet$ We use flat probability distributions as benchmarks for input parameters in the model. Although qualitative features seem to be insensitive to this assumption, inputting more stringy properties should allows us to consider more realistic probability distributions, which may lead to more precise statements.

$\bullet$ The model presented here provides a framework/skeleton for studying the naturalness issue. Surely one can build more features into the model to study high energy physics phenomenology and cosmology. For example, one can easily raise $M_{\susy}$ orders of magnitude by introducing another scalar field that undergoes SSB.

\section{Summary and Conclusion}\label{sec:summary}

Let us summarize the main result in this work. 
By combining the RKU model, $\overline{\rm D3}$-brane from the KKLT model and the Higgs sector in the SSM, we construct a model with statistically preferred exponentially small positive cosmological constant and desirable SUSY-breaking scale. EW scale also emerges from this model. Key observation is that there exists an upper bound (\ref{upperD2}) for $\mathcal{D}$ (\ref{3amigos}), which is sum of uplift contribution (on the vacuum energy) from $\overline{\rm D3}$-brane and the downward push contribution from the Higgs sector, and the upper bound is exponentially small. For phenomenological purpose, this simply implies $\mathcal{D}\simeq0$, that is the SUSY-breaking scale is closely tied to the SSB of the Higgs sector.

We adopt a statistical approach to explore the string theory landscape. In the patch of the landscape where a minimum solution exists, positive $\Lambda>0$ is tightly constrained and $\Lambda\propto e^{-2x}$ for large $x$ where the probability distribution $P(\Lambda) \propto \Lambda^{-1+k}$ ($1 >k> 0$) diverges when $\Lambda\sim 0^{+}$. Choosing a small $k$ so that the median $\Lambda_{50 \%}$ matches the observed $\Lambda_\text{obs}$, a minimum solution exists only when a new mass scale ${\bf m}$ takes value ${\bf m} = m_{\rm EW} \simeq  m_{\susy} \simeq 10^2$ GeV.

On the other hand, the model allow negative $\mathcal{D}<0$ in principle, which brings us AdS solution. When $\Lambda<0$, the allowed range for solution (i.e., $\Delta$ in eq.\eqref{Delta}) or eq.(\ref{minusD})) increases with $|\mathcal{D}|$. The value of $\Lambda$ is now dominated by the $\mathcal{D}$ factor and there is no exponential suppression $e^{-2x}$ in $\Lambda$, which means that there is no divergent behavior in $P(\Lambda)$. Sweeping through flux parameter space would give a smooth $P(\Lambda<0)$.  The key feature is depicted in the sketch of $P(\Lambda)$ in figure \ref{fig:proL}.

To conclude, one of the main attractions of string theory is that it offers a resolution to perturbative quantum gravity. Although some are disappointed of the existence of the string landscape, dashing any hope of ``uniqueness", and so of its predictive power, we opine that string landscape helps to ameliorate the naturalness problem. In this paper, we show that the string landscape can actually provide a different way to understand fundamental physics. 

In QFT, there are infinitely many possible solutions (i.e., local meta-stable vacua), and we have to fine-tune the parameters (after particle content and interactions are fixed) to yield the one that agrees with our observed universe. In string theory, in our framework, solutions are much more limited and we have to tune some parameters (i.e., choice of fluxes) to obtain any solution. Once we identify the very limited set of solutions, the various scales ($\Lambda$, $m_{\rm EW}$ and $m_{\susy}$) turn out to have the right order of magnitude to that in the universe we live in. In fact, within the brane inflationary scenario where an $\overline{\rm D3}$-brane is left over in the same warped throat as the SSM D3-branes, little tuning of parameters is needed to settle in the meta-stable vacuum state of our universe today. We believe this is an improvement in understanding nature. At the very least, this provides a new perspective to learn about our universe.

Hopefully, the approach adopted here will provide clues to find out where we are hiding in the string theory landscape. Centuries ago, (wo-)men tried to understand their place on earth. Last century, we learned our planet earth's place in the universe. Hopefully, applying naturalness as a guide, we'll learn more about our universe's place in the string landscape.

\acknowledgments

We thank Stefano Andriolo, Andy Cohen, Shing Yan Li, Tao Liu, Hoang Nhan Luu and Sam Wong for valuable comments.
This work is supported by the AOE grant AoE/P-404/18-6 issued by the Research Grants Council (RGC) of the Government of the Hong Kong SAR China.

\appendix

\section{$T$-Dependence of the ${\overline{\rm D3}}$-Brane Term and the Higgs Terms}
\label{appendix:barD3}

The form we use in the text for the $\overline{\rm D3}$-brane, $V_3\propto (2t)^{-n}$, where $n=3$ as applied in the KKLT scenario \cite{Kachru:2003aw}  or $n=2$ as adopted in reference \cite{Kachru:2003sx}. In the text, we allow both possibilities. Here is a clarification.
Let us start with the ten-dimensional metric for the four-dimensional spacetime and the  six-dimensional internal space,
\begin{equation}
ds^2=G_{MN}dX^MdX^N=e^{2A(y)-6u(x)}g_{\mu \nu}dx^{\mu} dx^{\nu} + e^{-2A(y) +2u(x)}{\hat g}_{mn}dy^mdy^n\;,
\end{equation}
where $e^{A(y)}$ is the warp factor in the six-dimensional internal space $X_6$ and $u(x)$ is the breathing mode for the variation of the size of the internal space as a function of the spacetime coordinate $x^{\mu}$. The factor $e^{-6u(x)}$ is a convenient choice for which the gravitational action in four dimensions will appear in the Einstein frame.
Here ${\hat g}_{mn}$ is a reference metric with the dimensionless internal volume $\cal V$
\begin{equation}
\int_{X_6} d^6 y \sqrt{{\hat g}} e^{-4A}={\cal V}\;,
\end{equation}
which is related to the K\"ahler modulus $T=t+i\tau$, 
\begin{equation}
g_s^2\left(\frac{M_\text{Pl}}{M_S}\right)^2={\cal V}=\left(T+{\bar T}\right)^{3/2} \gg 1\;,
\end{equation}
where $M_\text{Pl}$ is the reduced four-dimensional Planck mass, $g_s$ is the string coupling and $M_S$ is the string scale. Let the tension of an ${\overline{\rm D3}}$-brane be $T_3$. Sitting at $y=y_{0}$ yields a contribution to the four-dimensional stress-energy tensor given by
\begin{equation}
\hat{T}_3g_{\mu \nu}= T_3e^{8A-12u}g_{\mu \nu} \delta \left(y-y_{0}\right)\;.
\end{equation}
So its contribution to the effective potential is
\begin{equation}
V_3=\int_{X_6} d^6y \sqrt{{\hat g}} e^{-4A} p \hat{T}_3 = pT_3e^{4A(y_0) -12u(x)}\;,
\end{equation}
where we allow $p$ number of $\overline{\rm D3}$-branes, which are sitting in a warped throat. Since we are considering the single K\"ahler modulus case so $u(x)$ is the fluctuating mode of $T+ {\bar T}=2t$, we can display the K\"ahler modulus more explicitly in the metric, i.e.,  in terms of $T$, or  $e^u \to (2t)^{1/4}$, 
\begin{equation}
ds^2= (2t)^{-3/2}e^{2A(y)}\tilde{g}_{\mu \nu}dx^{\mu} dx^{\nu} + (2t)^{1/2} e^{-2A(y)}\hat{\tilde{g}}_{mn}dy^mdy^n\;.
\end{equation}
With this metric, one obtains \cite{Kachru:2003aw}
\begin{equation}
V_3= \frac{p T_3e^{4A(y_0)}}{(2t)^3}  \;.
\end{equation}
To look at $A(y_0)$, let us consider a Klebanov-Strassler throat \cite{Klebanov:2000hb}, where $K$ is the NS-NS flux and $M$ is the RR flux. If a stack of $p$ ${\overline{\rm D3}}$-branes sit at the bottom of such a throat, meta-stability requires \cite{Kachru:2002gs}
\begin{equation}
g_S^2 M^2 \gg p\;,
\end{equation}
which is easy to satisfy for large enough $M$. (In the text, we sometimes set $p=1$).
For $K \gg M \gg p$, we can get a very small warp factor 
\begin{equation}
e^{4A(y_0)} = {r_1} e^{-\frac{8\pi K}{3 g_s M}}\;.
\end{equation}
Here $r_1$ is the physical volume modulus of the throat, so it scales with the volume, i.e., $r_1 \propto t$, and  one has \cite{Kachru:2003sx,Kachru:2019dvo},
\begin{equation}
V_3 \simeq  \frac{pT_3 e^{-8\pi K/{3 g_s M}}}{(T+\bar{T})^2} = \frac{p\tilde{T}_3}{(T+\bar{T})^2}\;,
\label{neq2}
\end{equation}
where $\tilde{T}_3$ is the warped ${\overline{\rm D3}}$-brane tension. This $n=2$ form is generally adopted in the literature, which we shall follow. 

If the actual flux compactification is composed of multiple throats (and a bulk), so that a change in $r_1$ can be compensated by a change in $r_2$ of another throat while $t$ is unchanged; that is, $r_1(T,U_i)$ is actually a function of complex structure moduli as well \cite{Dudas:2019pls}. However, as $t$ increases, both $r_1(U_i)$ and $r_2(U_i)$ will scale with $t$, leading to eq.(\ref{neq2}).
Since the $U_i$ have been stabilized already in our model, its dependence on $U_i$ has been absorbed into the definition of $\tilde{T}_3$.


Since the SSM branes also live in a throat, the same argument applies to the Higgs terms $D_h$ and $S_h$ in eq.(\ref{3amigos}). So it is reasonable to choose $n_D=n_S=n=2$, i.e.,
\begin{equation}
\label{canonical}
V_3+D_h+S_h= \frac{p\tilde{T}_3}{(T+\bar{T})^{n}}+\frac{\hat{D}_h}{(T+\bar{T})^{n_D}}+\frac{\hat{S}_h}{(T+\bar{T})^{n_S}}=\frac{p\tilde{T}_3+ \hat{D}_h + \hat{S}_h}{(T+\bar{T})^2}\;,
\end{equation}
At times, we do consider the more general cases in the text.

\section{Single Uplift Model}\label{appendix:review}

We set the $M_\text{Pl}=1$ throughtout. The model we would like to consider is no-scale potential with single uplift contribution added and only one non-perturbative term $W_{\rm NP}=Ae^{-aT}$, in which scalar potential could be expressed as
\begin{align}
	\label{KUA3}
	V&\simeq \left(-\frac{a^3A\mathcal{W}}{2}\right)\lambda(x,y)\;,\nonumber\\
	\lambda(x,y)&= \rho e^{-2x}\frac{x+3}{x^2} -\frac{e^{-x}}{x^2}\cos{y}+\frac{Q}{x^n}, 
\end{align}
where $\rho=-A/(3\mathcal{W})>0$ and $x>0$. Different $Q$ corresponds to the three cases:
\begin{align}
	\text{no-scale}&:\;Q=0 \;,\nonumber\\
	{\rm KU}&:\;  Q=C=-\frac{3\xi a^{3/2} \mathcal{W}}{32\sqrt{2}A}, \quad n=9/2 \nonumber \\
	{\rm KKLT} &:\; Q=\tilde{D}_3=-\frac{\tilde{T}_3}{2aA\mathcal{W}}, \quad n=2
\end{align}
where $\mathcal{W}(U_i,S)$ is a real number with the complex structure moduli and the dilaton stabilized after flux compactifiction; and $x=at$ and $y=a\tau$ are real parameters. K\"ahler uplift (KU) model has been extensively studied \cite{Balasubramanian:2004uy,Westphal:2006tn,Rummel:2011cd,deAlwis:2011dp,Sumitomo:2012vx}. 
The $\xi$-term in the K\"ahler potential $K$ (\ref{Modele}) \cite{Becker:2002nn}
is a perturbative correction descending from an $\alpha'^3$ curvature correction in 10-dimensions. It arises in the 4-loop correction to the $\beta$-function of the world-sheet $\sigma$-model \cite{Gross:1986iv}.
It effectively uplifts the vacuum energy via the $F$-term potential already presented in eq.\eqref{eq:vf} and eq.\eqref{eq:kahler}, where we assume higher order terms in $\xi/\mathcal{V}$ in the KU model can be neglected. So the KU model reduces to the above form (\ref{KUA3}). The KKLT model \cite{Kachru:2003aw,Kachru:2003sx} contains the $\overline{\rm D3}$-brane tension $\tilde{T}_3$ which explicitly breaks SUSY.
 
Three cases have different physics, which nevertheless are related to each other. First, for the no-scale model with $Q=0$, the minimum of this potential is achieved by solving $\partial_x\lambda=\partial_y\lambda=0$,
\begin{equation}
	y=0\;,\quad \rho (2x+3)=e^{x}\;.
\end{equation}
One can easily check that above solution is stable because $\left.\partial_x^2\lambda\right|_\text{ext}>0$ and $\left.\partial_y^2\lambda\right|_\text{ext}>0$. This stable solution satisfies $\langle D_TW\rangle=0$, yielding a supersymmetric AdS vacuum, where
\begin{equation}
\label{susyv}
	\Lambda=V_\text{min}=-3 e^{K} |W|^2=-\frac{a^3 A^2 e^{-2x}}{6x}<0\;.
\end{equation}

Next, we turn on $Q$. If we assume 
\begin{equation}
	\rho e^{-2x}\frac{x+3}{x^2}\ll\frac{Q}{x^n}\;,
	\label{criteria}
\end{equation}
so that the $e^{-2x}$-term can be neglected, the potential is reduced to a two-term structure. A solution is obtained by balancing the last two terms in eq.(\ref{KUA3}), which is,
\begin{equation}
	y=0\;,\quad Q=\frac{1}{n}e^{-x}x^{n-2}(x+2)\;.
	\label{eq:2sln}
\end{equation}
If we treat $Q$ as an input parameter, then $x$ at a local minimum must satisfy eq.(\ref{eq:2sln}). To have a locally stable solution, second derivative of $\lambda$ at the extremum should be positive, 
\begin{equation}
\label{xconstraint}
\left.\partial_x^2\lambda\right|_{\rm ext} = \frac{e^{-x}}{x^3}\left[n\left(1 +\frac{2}{x}\right) -x\left(1 +\frac{3}{x} +\frac{4}{x^2}\right)\right] >0
	\end{equation}
or $n>1+x+{2}/{(x+2)}>2$;
this says that there is no solution for the $Q=\tilde{D}_3$ case (with $n=2$).

 In the $Q=C$ case, where $n=9/2$, the two-term structure has been studied \cite{Rummel:2011cd}. Here, $x (C)$ and $C$ are  bounded: 
\begin{equation}
\label{xmax}
x <x_{\rm max}=\left(3 +\sqrt{89}\right)/4 \simeq 3.109, \quad \quad C< C_{\rm max}\simeq 0.865
\end{equation}
Here $x=5/2$ (or equivalently $C=0.811$) yields a Minkowski solution. For a slightly bigger (smaller) $x$ (or the corresponding $C$ (\ref{eq:2sln})), the solution is a dS (AdS) vacuum.
The dS ($\lambda_\text{ext} >0$) vacuum solution is only meta-stable since $\lambda(x\to\infty)=0$. As we scan over the parameter of the model, namely $C$, to go from AdS to dS, $\lambda=0$ does not hold a special place. In other word, $P(\Lambda)$ should be smooth around $\Lambda=0$. 

Now, let us check the validity of the assumption (\ref{criteria}). 
Including the $e^{-2x}$-term in $V$ (\ref{KUA3}), $\tilde{D}_3$ and $\rho$ are independent parameters, so the assumption can easily be avoided.  Keeping the  $e^{-2x}$-term allows dS vacuum solution in the KKLT model \{cf.\cite{Choi:2005ge}\};  but there is no special preference for any particular value for $\Lambda$ (or $\lambda$) as $\rho$ and $\tilde{D}_3$ vary.

In the KU model, the situation is entirely different, where $\rho=\sigma/C$ and $\sigma= {\xi a^{3/2}}/{32\sqrt{2}}$.
Then at the extremum $\partial_x\lambda=0$, we have
\begin{equation}
\label{quadratic}
\sigma=\frac{2}{9}x^{5/2} e^{-2x}(x+2)\rho \left(e^{x}-\rho(2x+3)\right)\;.
\end{equation}
This is a quadratic function, indicating two solutions of $\rho(x)$ in the $Q=C$ case.  Given the fact that $\sigma\sim 10^{-4}$ to $10^{-3}$ is a small number, those solutions could be approximated as
\begin{equation}
\label{2sol}
	\rho_{+}\simeq\frac{e^{x}}{2x+3}\;,\quad \rho_{-}\simeq \frac{9\sigma e^{x}}{2x^{5/2}(x+2)}\;.
\end{equation}
It is apparent that $\rho_{+}\gg\rho_{-}$, and they lead to different scenarios:\\
(1) $\rho_+$ indicates the opposite of eq.\eqref{criteria} and it satisfies the $\left.\partial_x^2\lambda\right|_\text{ext}>0$ condition for large $x$, which yields a negative $\lambda_{\rm ext}$. That is, the uplift $Q=C=\sigma/\rho_{+}$ is too small to lift the vacuum to a dS solution. Here, $C$ simply plays the role of a correction term to the above AdS solution case (\ref{susyv}).\\
(2) $\rho_-$ satisfies criteria eq.\eqref{criteria}. The potential reduces to the two-term structure as before, where the $e^{-2x}$-term plays the role of a correction only, shifting $x_{\rm max}$ (\ref{xmax}) slightly as
\begin{equation}
x_{\rm max}'=x_{\rm max}+\epsilon(\sigma)\;,\quad \epsilon(\sigma)\simeq 4.095\sigma-2.649\sigma^2+\cdots\;,
	\label{xcorrection}
\end{equation}
which is a small correction for small $\sigma \sim 10^{-3}$.
So the KU model has a dS solution only for $2.5< x <3.12$ (with a corresponding value for $C$ or $\rho_-$).
 
To investigate the SUSY-breaking scale of KU model, we expand $\langle D_T W\rangle$ in power series of $\xi/\mathcal{V}$ as
\begin{equation}
	\langle D_T W\rangle ={\langle D_T W\rangle}^{(0)}+\frac{\xi}{\mathcal{V}}{\langle D_T W\rangle}^{(1)}+\cdots\;.
\end{equation}
After balancing $e^{-x}$, $\rho e^{-2x}$ and $C$-term in the potential, we plug the solution into this quantity, up to first order in $\xi/\mathcal{V}$, and we arrive at
\begin{equation}
	\langle D_T W\rangle \simeq \frac{3a^{5/2}A\xi}{128\sqrt{2}}\left|f(\rho,x)\right|\;,
\end{equation}
where function $f$ is
\begin{equation}
	f(\rho,x)=\frac{e^{-x}}{3x^{7/2}}\left[48x-16x \frac{e^x}{\rho}+\frac{9e^{2x}}{(x+2)\rho^2}\right]\;.
\end{equation}
Apparantly, supersymmetry is preserved when $\xi \to 0$. However, the presence of K\"ahler uplift does not guarantee SUSY-breaking stable solution. As long as $f=0$ and solution is stable, it is a supersymmetric vacua.

\section{Racetrack Model with Single Uplift}
\label{appendix:racetrack}

The racetrack model considers multiple non-perterbative terms in the superpotential and here we take $W_{\rm NP}=Ae^{-aT}+Be^{-bT}$ for simplicity. Scalar potential with only one uplift contribution can be expressed as
\begin{align}
	V(T)&\simeq\left(-\frac{a^3A\mathcal{W}}{2}\right)\lambda(x,y)\nonumber\\
	\lambda(x,y)&=\frac{\rho}{x^2}\left[e^{-2x}(x+3)+\frac{\beta}{z^2}e^{-2\beta x}(\beta x+3)+\frac{1}{z}e^{-(\beta+1)x}\cos{\left[(1-\beta)y\right]}(2\beta x+3\beta+3)\right]\nonumber\\
	&\qquad-\frac{e^{-x}}{x^2}\cos{y}-\frac{\beta}{z}\frac{e^{-\beta x}}{x^2}\cos{(\beta y)}+\frac{Q}{x^n}\;,
	\label{double}
\end{align}
where $Q$ again corresponds to the three cases, including the racetrack KKLT (RKKLT) model,
\begin{align}
	\text{no-scale}&:\;Q=0 \;,\nonumber\\
	{\rm RKU}&:\;  Q=C=-\frac{3\xi a^{3/2} \mathcal{W}}{32\sqrt{2}A}, \quad n=9/2 \nonumber \\
	{\rm RKKLT} &:\; Q=\tilde{D}_3=-\frac{\tilde{T}_3}{2aA\mathcal{W}}, \quad n=2
\end{align}
Solution to racetrack no-scale model is given by solving $\partial_x\lambda=\partial_y\lambda=0$,
\begin{equation}
	y=0\;,\quad \rho=\frac{z e^{(\beta+1)x}}{e^x (2 \beta  x+3)+ z
		e^{\beta x}(2 x+3)}\;.
	\label{racetrack ads}
\end{equation}
There exist a hidden constraint for parameter $z<0$ from the fact that $\rho>0$, which is
\begin{equation}
	z<-e^{(1-\beta)x}\frac{2\beta x+3}{2x+3}\;.
	\label{upz}
\end{equation}
One can easily check that $ \langle D_T W\rangle=0$ for this solution \eqref{racetrack ads}, indicating that this is a supersymmetric AdS solution. 

Similar to appendix \ref{appendix:review}, when considering the RKU model, we reduce the potential to a less complicated structure given the fact that $\sigma=\xi a^{3/2}/32\sqrt{2}$ is a small number. The extremum condition  $\partial_x\lambda=0$ again yields a quadratic equation for $\rho$: the two solutions $\rho_\pm(x,z,\beta)$ lead to different scenarios. $\rho_{+}$ makes the $C$-term a small correction to above no-scale AdS solution and $\rho_{-}$ enables us to safely neglect terms involving $e^{-2x}$, $e^{-2\beta x}$ and $e^{-(1+\beta)x}$, leaving the potential with a simpler structure,
\begin{equation}
	\lambda(x,y)\simeq -\frac{e^{-x}}{x^2}\cos{y}-\frac{\beta}{z}\frac{e^{-\beta x}}{x^2}\cos{(\beta y)}+\frac{Q}{x^{n}}\;.
	\label{racetrackmodel}
\end{equation}
Due to the fact that $z<0$ and $\beta$ close to $1$, the second term in this potential is positive, which effectively provides an uplift contribution to cancel part of the first negative $e^{-x}$-term. Same as in last appendix, $\left.\partial_x^2\lambda\right|_{\rm ext}>0$ together with constraint eq.\eqref{upz} gives, in the large $x$ approximation,
\begin{equation}
\left.\partial_x^2\lambda\right|_{\rm ext} \simeq \frac{e^{-x}}{x^3}[n-x] +\frac{\beta^2}{z}\frac{e^{-\beta x}}{x^3} [n-\beta x] > 0 \;,
\end{equation}
which allows a large $x$ solution to exist, provided $\beta>1$ is close to one. Note that the upper bound on $x$ in the single term case (\ref{xconstraint}) disappears for appropriate $\beta \gtrsim 1$ here. Since $\lambda \sim e^{-x}$ so $\Lambda \propto e^{-2x}$ and large $x$ is allowed, the racetrack now offers the possibility of an exponentially small $\Lambda$. It is clear that there are some constraints on the input parameters for a large $x$ solution to exist.  As we see below, these constraints on the racetrack play another important role.

To investigate the constraints on $\lambda_{\rm ext}$ we express the input parameter $z$ and $Q$ in terms of the outputs $\lambda_{\rm ext}$ and $x$,
\begin{align}
	\frac{1}{z}&=\frac{e^{\beta x}}{\beta(\beta x+2-n)}\left[-e^{-x}(x+2-n)+nx^2\lambda_{\rm ext}\right]\;,\nonumber\\
	Q&=\frac{x^{n-1}}{\beta x+2-n}\left[(\beta-1)e^{-x}+(\beta x^2+2x)\lambda_{\rm ext}\right]\;.
\end{align}
Hessian at the extremum ($\partial_y\lambda=0$ yields $y=0$) is the key for the existence of locally stable solution. Here we express them in $(x,\beta,\lambda_{\rm ext})$. One can easily get $\partial_x\partial_y\lambda|_\text{ext}=\partial_y\partial_x\lambda|_\text{ext}=0$ and 
\begin{align}
	\left.\partial_x^2\lambda\right|_\text{ext}&=e^{-x}\left(\frac{\beta-1}{x^2}-\frac{(n-2)(\beta+1)}{x^3}+\cdots\right)-\lambda_{\rm ext}\left(\frac{n\beta}{x}+\frac{n}{x^2}+\cdots\right)\;,\nonumber\\
	\left.\partial_y^2\lambda\right|_\text{ext}&=-e^{-x}\left(\frac{\beta-1}{x^2}-\frac{(n-2)(\beta-1)}{x^3}+\cdots\right)+\lambda_{\rm ext}\left(\frac{n\beta}{x}+\frac{n(n-2)}{x^2}+\cdots\right)\;,
\end{align}
where we have expanded in large $x$ and replace $z$ and $Q$ by $\lambda_{\rm ext}$ and the value of $x$ there. Requiring both $m_x^2\equiv\left.\partial_x^2\lambda\right|_\text{ext}$ and $m_y^2\equiv\left.\partial_y^2\lambda\right|_\text{ext}$ to be semi-positive gives us
\begin{equation}
	\lambda_\text{min}\leq\lambda_\text{ext}\leq\lambda_\text{max}\;,
\end{equation}
where $\lambda_\text{max}$ comes from $m_x^2 \geq 0$ while $\lambda_\text{min}$ comes from $m_y^2 \geq0$. 

For RKU model, $Q=C$ and $n=9/2$. Hessian at local minimum solution gives \cite{Sumitomo:2013vla}
\begin{equation}
	m_x^2\geq0\;,\quad m_y^2=-m_x^2+\lambda_{\rm ext}\left(\frac{27}{4x^2}\right)+\cdots\geq0\;,
\end{equation}
implying $\lambda_{\rm ext}>0$. Here the constraints on $\lambda_{\rm ext}$ can be easily found, for large $x$,
\begin{equation}
	e^{-x}\frac{2(\beta-1)}{9\beta x}\left(1-\frac{5(\beta+1)}{2\beta x}+\cdots\right)\lesssim\lambda_\text{ext}\lesssim e^{-x}\frac{2(\beta-1)}{9\beta x}\left(1-\frac{5(\beta+1)}{2\beta x}+\frac{3}{\beta x}+\cdots\right)\;.
\end{equation}
For large $x$ and $\beta>1$, the lower bound $\lambda_{\rm min}$ is positive, which ensures a dS solution. 
In some sense, racetrack allows us to trade the bound on $x$ (\ref{xmax}) for the bounds on $\lambda$ (or equivalently the bounds on $\Lambda$). Again, $\lambda_\text{min} \to \lambda_\text{max}$ as $x \to \infty$. So, to leading order, we obtain
\begin{equation}
	\lambda_\text{ext}\simeq e^{-x}\frac{2(\beta-1)}{9\beta x}\;,\quad C=-\frac{3\xi a^{3/2} \mathcal{W}}{32\sqrt{2}A}\simeq e^{-x}\frac{2(\beta-1)}{9\beta} x^{7/2}\;.
\end{equation}
Under this approximation, the cosmological constant is obtained
\begin{equation}
	\Lambda\simeq\frac{64\sqrt{2}a^{3/2}A^2(\beta-1)^2}{243\beta^2\xi}e^{-2x}x^{5/2}\simeq\frac{3\xi {\mathcal{W}}^2}{4(2t)^{9/2}}\;.
\end{equation}
The racetrack model introduces parameter $\beta$, which renders non-trivial the axionic direction and gives a lower bound on $\lambda_\text{ext}$. With the uplift $\xi$-term, the $e^{-2x}$ factor suggests that the positive $\Lambda$ can be naturally exponentially small. Here the masses of the $U_i +\bar{U}_i$ and their axionic modes are very light \cite{Tye:2016jzi}.
Unfortunately, $m^2_{3/2}  \simeq 4 \mathcal{V} \Lambda/(3\xi) \sim 10^{-110} M_{\rm Pl}^2$, which is far too small to be relevant for phenomenology.

In RKKLT model $Q=\tilde{D}_3$ and $n=2$. If the $\rho$-related terms in the potential (\ref{double}) are dropped, we cannot find a solution. The reason is obvious. Following the same analysis, the bound for $\lambda_{\rm ext}$ is
\begin{align}
	e^{-x}&{(\beta-1) \over n \beta x} \left(1 - {(n-2) (\beta +1) \over \beta x}\right) \leq \lambda_{\rm ext} \nonumber\\
	&\lesssim e^{-x}{(\beta-1) \over n \beta x}  \left(1 - {(n-2)(\beta + 1) \over  \beta x } + {(n-3)\over \beta x } +\frac{2(n-2)\beta +n-1}{\beta^2 x^2} +\cdots \right)\;,
	\label{leading lambda}
\end{align}
which is only valid when $n\geq3$. Including $\rho$-related term in potential can give us a solution; but it is arbitrary since the uplift contribution from tension of $\overline{\rm D3}$-brane is arbitrary.

\section{Racetrack Model with Combined Uplift}\label{appendix:combined}

In the RKU model we can neglect the $\rho$-related terms and simplify the potential structure, while we may or may not do so in the RKKLT model, as $\rho$ is a relatively free parameter. If we combine them together, we can safely simplify the potential structure and have nice property. The potential function we are interested in is $V(T)$
as defined in eq.\eqref{eq:model} and here we keep the general $n$ to discuss the influence of different choice of $n$. We have four degrees of freedom in this classical potential after stabilization. We can express every quantity as functions of $(x,\beta,\lambda_{\rm ext},\tilde{D})$, or as functions of $(x,\beta,\lambda_{\rm ext},C)$. The former approach is adopted in the main text.  As an alternative, let us consider the later approach, i.e., we eliminate $\tilde{D}$ instead of $C$ here.

Parameter $C$ is well-defined and is positive for our consideration of geometry. The positivity of $\tilde{D}$ is undetermined due to the unknown detail in the SUSY breaking mechanism. We follow the similiar procedure and found bounds for local minimum solution, $\lambda_{\rm min}\leq\lambda_{\rm ext}\leq \lambda_{\rm max}$. Expressed in $(x,\beta,C)$, they are, for large $x$ and $\beta \gtrsim 1$,
\begin{align}
	\lambda_{\rm max}&\simeq e^{-x}\frac{\beta-1}{n\beta x}\left(1-\frac{(n-2)(\beta+1)}{\beta x}+\frac{n-3}{\beta x}+\cdots\right)-\frac{C(9-2n)}{2nx^{9/2}}\left(1-\frac{9}{2\beta x}+\cdots\right)\nonumber\\
	\lambda_{\rm min}&=e^{-x}\frac{\beta-1}{n\beta x}\left(1-\frac{(n-2)(\beta+1)}{\beta x}\right)-\frac{C(9-2n)}{2n x^{9/2}}\;.
\end{align}
The requirement $\lambda_{\rm max}\geq\lambda_{\rm min}$ is automatically satisfied for $n=9/2$, the case for the RKU model.
It gives a lower bound for $C$, when $n<9/2$,
\begin{equation}
		C\gtrsim\frac{4(3-n)}{9(9-2n)}x^{7/2}e^{-x}\frac{\beta-1}{\beta}\;.
\end{equation}
If $n=3$, $C>0$, which is trivial. If $n=2$, the solution still could exist but the value of $x$ is bounded by parameter $C$ and $\tilde{D}$.
One can also express them as functions of $(x,\beta,\tilde{D})$, and get the upper bound for $\tilde{D}$ as in eq.\eqref{eq:upDi}.

\section{Explicit Calculation and Approximation}\label{appendix:cal}

The general model is presented here. Scalar potential is given by
\begin{align}
	V(T)&\simeq \left(-\frac{a^3 A\mathcal{W}\mathcal{N}}{2}\right)\lambda(x,y)\;,\nonumber\\
	\lambda(x,y)&=-\frac{e^{-x}}{x^2}\cos{y}-\frac{\beta}{z}\frac{e^{-\beta x}}{x^2}\cos{(\beta y)}+\frac{C}{x^{9/2}}+\sum_i^M\frac{\tilde{D}_i}{x^{n_i}}\;,\nonumber\\
	C&=-\frac{3\xi a^{3/2} \mathcal{W}}{32\sqrt{2}A}\;,\quad \tilde{D}_i=\left(-\frac{2}{a^3A \mathcal{W}\mathcal{N}}\right)\left(\frac{a}{2}\right)^{n_i}\mathcal{D}_i\;.\label{eq:gmodel}
\end{align}
Here we keep the general $n_i$ and $i=1,2,\cdots, M$, indicating there could be more contributions other than K\"ahler uplift (i.e., $C$),  $\overline{\rm D3}$-brane and the Higgs sector (where $M=1$ and $n_1=n=2$).  Solving the extremum equation $\partial_x\lambda=\partial_y\lambda=0$ gives
\begin{equation}
	y=0\;,\quad \frac{\beta}{z}=\frac{e^{\beta x}}{\beta x+2}\left[-(x+2)e^{-x}+\frac{9C}{2x^{5/2}}+\sum_i\frac{n_i \tilde{D}_i}{x^{n_i-2}}\right]\;,
\end{equation}
which gives solution $\lambda_\text{ext}(z,\beta,C,\tilde{D}_i)$. There are $3+M$ degrees of freedom in this system. Parameter $z$ and $C$ are expressed as functions of $(x,\beta,\tilde{D}_i,\lambda_{\rm ext})$.
\begin{align}
	\frac{1}{z}&=\frac{e^{\beta x}}{\beta (2\beta x-5)}\left[-(2x-5)e^{-x}+9x^2\lambda_\text{ext}-\sum_i\frac{9-2n_i}{x^{n_i-2}}\tilde{D}_i\right]\nonumber\\
	C&=\frac{2x^{5/2}}{2\beta x-5}\left[(\beta-1)xe^{-x}+(\beta x+2)x^2\lambda_\text{ext}-\sum_i\frac{\beta x+2-n_i}{x^{n_i-2}}\tilde{D}_i\right]
\end{align}
Hessian of function $\lambda$ at the extremum solution is calculated step by step and expressed as functions of $(x,\beta,\tilde{D}_i,\lambda_{\rm ext})$ below.
\begin{align}
	\left.\partial_x^2\lambda\right|_\text{ext}&=-\frac{e^{-x}}{x^2}\frac{x^2+4x+6}{x^2}-\frac{\beta}{z}\frac{e^{-\beta x}}{x^2}\frac{\beta^2 x^2 +4\beta x+6}{x^2}+\frac{99 C}{4 x^{13/2}}+\sum_i\frac{n_i(n_i+1) \tilde{D}_i}{x^{n_i+2}}\nonumber\\
	&=\frac{9(2\beta^2x^2-3\beta x-10)}{2x^2(2\beta x-5)}\bigg[e^{-x}\frac{2(\beta-1)}{9\beta x}\frac{4\beta^2 x^2-10\beta x(\beta+1)+35\beta}{4\beta^2x^2-6\beta x-20}\nonumber\\
	&\qquad\qquad\qquad+\sum_i\frac{(9-2n_i)\tilde{D}_i}{9x^{n_i}}\frac{2\beta^2x^2-(3+2n_i)\beta x+5(n_i-2)}{2\beta^2x^2-3\beta x-10}-\lambda_\text{ext}\bigg]\nonumber\\
	&\simeq\frac{9(2\beta^2x^2-3\beta x-10)}{2x^2(2\beta x-5)}\bigg[e^{-x}\frac{2(\beta-1)}{9\beta x}\left(1-\frac{5(\beta+1)}{2\beta x}+\frac{3}{2\beta x}+\cdots\right)\nonumber\\
	&\qquad\qquad\qquad+\sum_i\frac{\tilde{D}_i}{x^{n_i}}\left(1-\frac{2n_i}{9}\right)\left(1-\frac{n_i}{\beta x}+\cdots\right)-\lambda_\text{ext}\bigg]\;,
\end{align}
\begin{align}
	\left.\partial_y^2\lambda\right|_\text{ext}&=\frac{e^{-x}}{x^2}+\frac{\beta^3}{z}\frac{e^{-\beta x}}{x^2}\nonumber\\
	&=\frac{9\beta^2}{2\beta x-5}\left[-e^{-x}\frac{2(\beta-1)}{9\beta x}\left(1-\frac{5(\beta+1)}{2\beta x}\right)-\sum_i\frac{\tilde{D}_i}{x^{n_i}}\left(1-\frac{2n_i}{9}\right)+\lambda_\text{ext}\right]\;,
\end{align}
and obviously $\left.\partial_x\partial_y\lambda\right|_\text{ext}=\left.\partial_y\partial_x\lambda\right|_\text{ext}=0$. Stability condition requires $m_x^2=\left.\partial_x^2\lambda\right|_\text{ext}\geq0$ and $m_y^2=\left.\partial_y^2\lambda\right|_\text{ext}\geq0$, which lead to
\begin{equation}
	\lambda_\text{min}\leq\lambda_\text{ext}\leq\lambda_\text{max}\;,
\end{equation}
where
\begin{align}
	\lambda_\text{max}&=e^{-x}\frac{2(\beta-1)}{9\beta x}\frac{4\beta^2x^2-10\beta x(\beta+1)+35\beta}{4\beta^2 x^2-6\beta x-20}\nonumber\\
	&\qquad\qquad\qquad\qquad+\sum_i\frac{(9-2n_i)\tilde{D}_i}{9x^{n_i}}\frac{2\beta^2x^2-(3+2n_i)\beta x+5(n_i-2)}{2\beta^2 x^2-3\beta x-10}\nonumber\\
	&\simeq e^{-x}\frac{2(\beta-1)}{9\beta x}\left(1-\frac{5(\beta+1)}{2\beta x}+\frac{3}{2\beta x}+\cdots\right)+\sum_i\frac{\tilde{D}_i}{x^{n_i}}\left(1-\frac{2n_i}{9}\right)\left(1-\frac{n_i}{\beta x}+\cdots\right)\;,\nonumber\\
	\lambda_\text{min}&=e^{-x}\frac{2(\beta-1)}{9\beta x}\left(1-\frac{5(\beta+1)}{2\beta x}\right)+\sum_i\frac{\tilde{D}_i}{x^{n_i}}\left(1-\frac{2n_i}{9}\right)\;.
\end{align}
The lower bound $\lambda_\text{min}$ is exact while the expression for $\lambda_\text{max}$ is for large $x$. Their difference  happens at higher order in $1/x$. Note that any $\tilde{D}_i$-term with $n_i=9/2$ vanishes here. This is expected since $9/2$ is precisely the power for the $C$-term in $V(T)$ (\ref{eq:gmodel}), so it would have been combined with $C$ in  $V(T)$ (\ref{eq:gmodel}) and not show up here.

To guarantee the existence of a solution, one requires $\lambda_{\rm max}\geq \lambda_{\rm min}$. This would give the result eq.\eqref{eq:upDi}, which is already presented in the main text. It is useful to make the approximation that
\begin{equation}
	\lambda_\text{ext}=f\lambda_\text{max}+(1-f)\lambda_\text{min}\;,
\end{equation}
where $0\leq f\leq 1$ is a real number interpolating between the two limits. By extracting the leading-order information of $e^{-x}$-term and $\tilde{D}_i$-term, we can write that
\begin{equation}
	\lambda_\text{ext}=e^{-x}\frac{2(\beta-1)}{9\beta x}\left(1-\frac{5(\beta+1)}{2\beta x}\right)\epsilon_1(x,f)+\sum_i\frac{\tilde{D}_i}{x^{n_i}}\left(1-\frac{2n_i}{9}\right)\epsilon_{2,i}(x,f)\;,
\end{equation}
where $\epsilon_1$ and $\epsilon_{2,i}$ are functions close to 1 for any $f$ and $x\gg1$.
\begin{align}
	\epsilon_1(x,f)&=\frac{2\beta x\left(4\beta^2x^2-10\beta x(\beta+1)+35\beta\right)}{\left(2\beta x-5(\beta+1)\right)\left(4\beta^2x^2-6\beta x-20\right)}f+(1-f)\nonumber\\
	\epsilon_2(x,f)&=\frac{2\beta^2x^2-(3+2n_i)\beta x+5(n_i-2)}{2\beta^2 x^2-3\beta x-10}f+1-f\;.
\end{align}
This reduces one degree of freedom in all quantities, which are now functions of $(x,\beta,\tilde{D}_i)$. The explicit expressions for parameter $z$ and $C$ after making the approxiamtion for $\lambda_\text{ext}$ are
\begin{align}
	C&\simeq \frac{2x^{3/2}g(x)}{3}\epsilon_3(x,f) - \sum_i\frac{2x^{9/2}}{9}\frac{n_i \tilde{D}_i}{x^{n_i}}\epsilon_{4,i}(x,f)\;,\nonumber\\
\frac{1}{z}& \simeq - \frac{e^{-(1-\beta)x}}{\beta^3}\epsilon_5(x,f)+\sum_i\frac{e^{\beta x}}{\beta^2}\frac{n_i\tilde{D}_i}{2x^{n_i-1}}\epsilon_{6,i}(x,f)\;,
\end{align}
where
\begin{align}
	g(x)&=x^2e^{-x}\frac{\beta-1}{3\beta}\;,\nonumber\\
	\epsilon_3(x,f)&=\frac{4\beta^2 x^2+2\beta x(9\beta /\epsilon_1-5\beta -1)-20(\beta+1)}{4\beta^2 x^2-10\beta x}\epsilon_1\;,\nonumber\\
	\epsilon_{4,i}(x,f)&=\frac{2\beta x n_i \epsilon_{2,i}+9(\beta x+2)(1-\epsilon_{2,i})-(9-4\epsilon_{2,i})n_i}{(2\beta x-5)n_i}\;,\nonumber\\
	\epsilon_5(x,f)&=\frac{2\beta x(1-2\beta(1-1/\epsilon_1))+5\beta^2(1-1/\epsilon_1)-5}{2\beta x-5}\epsilon_1\;,\nonumber\\
	\epsilon_{6,i}(x,f)&=\frac{2\beta x-9\beta x/n_i}{2\beta x-5}(1-\epsilon_2)\;.
\end{align}
It is easy to see that for $f=0$, all expression could be simplified. Especially, the dependence of $\tilde{D}_i$ in $1/z$ would disappear, which is good for statistical analysis;  this is the reason we take this approximation in the main text. The full expression for $\mathcal{W}$ is
\begin{equation}
	\mathcal{W}\simeq-\frac{32 A}{9\xi} \left(\frac{2x}{a}\right)^{3/2}\left[g(x) \epsilon_3-\frac{x^3}{3}\sum_i\frac{n_i\tilde{D}_i}{x^{n_i}}\epsilon_{4,i}\right]\;.
\end{equation}
Along with
\begin{equation}
	\lambda_\text{ext}\simeq\frac{2}{3x^3}g(x)\left(1-\frac{5(\beta+1)}{2\beta x}\right)\epsilon_1+\sum_i\frac{\tilde{D}_i}{x^{n_i}}\left(1-\frac{2n_i}{9}\right)\epsilon_{2,i}\;,
\end{equation}
$\Lambda(x,\beta,\tilde{D})$ becomes
\begin{align}
	\Lambda&=V_\text{min}\simeq\left(-\frac{a^3 A \mathcal{W}\mathcal{N} }{2}\right)\lambda_\text{ext}\nonumber\\
	&\simeq \frac{128A^2\mathcal{N}}{27\xi}\left(\frac{a}{2x}\right)^{3/2}\left[g(x)\epsilon_3-\frac{x^3}{3}\sum_i\frac{n_i\tilde{D}_i}{x^{n_i}}\epsilon_{4,i}\right]\times\nonumber\\
	&\qquad\qquad\qquad\left[2g(x)\left(1-\frac{5(\beta+1)}{2\beta x}\right)\epsilon_1+3x^3\sum_i\frac{\tilde{D}_i}{x^{n_i}}\left(1-\frac{2n_i}{9}\right)\epsilon_{2,i}\right]\;,
	\label{eq:generalL}
\end{align}
Rearrange a little and we obtain the expression
\begin{align}
	\Lambda&\simeq \frac{3\mathcal{N}\xi \mathcal{W}^2}{4(2t)^{9/2}}\left(1-\frac{5(\beta+1)}{2\beta x}\right)\frac{\epsilon_1}{\epsilon_3}+\sum_i\frac{\mathcal{D}_i}{(2t)^{n_i}}\left[\left(1-\frac{2n_i}{9}\right)\epsilon_{2,i}+\left(1-\frac{5(\beta+1)}{2\beta x}\right)\frac{2n_i}{9}\frac{\epsilon_1\epsilon_{4,i}}{\epsilon_3}\right]\nonumber\\
	&\overset{f\to 0}{\simeq} \frac{3\mathcal{N}\xi \mathcal{W}^2}{4(2t)^{9/2}}\left(1-\frac{9(\beta+1)}{2\beta x}+\cdots\right)+\sum_i\frac{\mathcal{D}_i}{(2t)^{n_i}}\left(1-\frac{n_i(\beta+1)}{\beta x}+\cdots\right)\;.
	\label{eq:generalLL}
\end{align}
The upper bound for other contributions $\mathcal{D}_i$ is,
\begin{equation}
	\sum_i\frac{n_i\mathcal{D}_i}{(2t)^{n_i}}\left[1-\frac{2n_i}{9}-\frac{\epsilon_{4,i}}{3\epsilon_3}\right]\lesssim \frac{9\mathcal{N}\xi \mathcal{W}^2}{8(2t)^{9/2}}\frac{1}{\epsilon_3}\;.
	\label{eq:upd}
\end{equation}
To be specific, only if the quantity in the square bracket is positive, this upper bound is meaningful.

\section{Justification of Dropping the $e^{-2x}$-term}\label{appendix:drop}

Let us check the validity in dropping the $e^{-2x}$ and other doubly suppressed terms in potential $V_F$. The full expression for $V(T)$ \eqref{eq:model} contains the additional doubly suppressed terms, as given in eq.\eqref{double}. Here the issue is insensitive to the racetrack so let us simplify the discussion by considering the case of combined uplift without the racetrack.

Since $C=\sigma/\rho$ and $\tilde{D}\propto \rho $,  we obtain a quadratic equation for $\rho>0$ at the extremum $\partial_x\lambda=\partial_y\lambda=0$, similar to eq.\eqref{quadratic}, which is
\begin{equation}
	\left(\alpha+\delta\right)\rho^2-\gamma \rho+\sigma=0\;,
\end{equation}
where $\alpha(x)\sim e^{-2x}$ and $\gamma(x)\sim e^{-x}$ are functions of stablized $x$ only and they are positive; $\delta$ is proportional to $\tilde{D}$ and it can be treated as a free parameter here. There are two solutions for the above equation: $\rho_{+}$ and $\rho_{-}$, similar to eq.(\ref{2sol}). As long as $\delta>-\alpha$, both solutions are positive, $\rho_\pm>0$. 
Given the fact that $\sigma\sim10^{-3}$ and for a not too large $\delta$, $\rho_{+}\gg\rho_{-}$, where $\rho_{+}$ would lead to AdS solutions like in the KU model. Again, only the $\rho_{-}$ solution can yield dS vacua, in which case the doubly suppressed terms ($e^{-2x}$, $e^{-(\beta +1)x}$ and $e^{-2\beta x}$) are small compared to the singly-suppressed terms and the $\tilde{D}$ terms. The correction from them is expected to be small, as indicated by the small shift in $x_{\rm max}$ (\ref{xcorrection}). If $\delta<-\alpha$, $\rho_{+}$ would flip sign and there is only one physical solution, $\rho_{-}>0$. But for the dominating negative $\tilde{D}$, only AdS solutions are available. There is no exponential suppression constraint 
so $P(\Lambda)$ is smooth for $\Lambda <0$, as shown in figure \ref{fig:proL}. In seeking dS meta-stable solutions, neglecting the doubly suppressed terms is a good approximation.

Now turn to racetrack model with combined uplift, which is the situation in main text. If these terms in the potential is negligible after stabilization of $x$, they are reasonable to be neglected from the beginning. Typically such terms have the structure as
\begin{equation}
	\Delta V_F \sim \sum_{i,j} a^3\mathcal{N} \beta_i \beta_j A_i A_j \frac{e^{-(\beta_i+\beta_j)x}}{6 x}\left(1+\mathcal{O}\left(1/x\right)\right)\propto \mathcal{N}\frac{e^{-2x}}{x}\;,
\end{equation}
where $i,j$ stand for different non-perturbative terms, that is, $\beta_i=1$ or $\beta_i=\beta \gtrsim 1$. At the minimum, the $\Lambda =V_F$ can be written as 
\begin{equation}
	\Lambda \simeq \frac{256A^2\mathcal{N}}{27\xi}\left(\frac{a}{2x}\right)^{3/2}g(x)^2\epsilon_1\epsilon_3+\cdots\propto \frac{\mathcal{N}}{\xi}x^{5/2}e^{-2x}\;.
\end{equation}
$\epsilon_1$ and $\epsilon_3$ are the small real numbers close to 1. As indicated by power of $x$ in the above terms, we see that original $e^{-2x}$-terms contribution in $V_F$ is small compared to $\Lambda$ for $x\sim \mathcal{O}(100)$ and $\xi \sim 10^{-2}$. It is safe to neglect them in the first place.

\section{Note on $\mathcal{N}(U_i, S)$}
\label{appendix:prefactor}

In the simplified model, we assume that the dilaton $S$ and the complex structure moduli $U_i$ have been stabilized already \cite{Giddings:2001yu}. Their stabilization introduces a constant $W_0(U_i,S)$ in the superpotential $W$. Since they are also present in the K\"ahler potential $K$, they also contribute an overall factor $\mathcal{N}(U_i, S)$ to the $F$-term $V_F$ in potential $V$. Here we like to commend on this. 
First, it has been shown \cite{Kachru:2019dvo} that the ten-dimensional stress-energy reduces to the four-dimensional $V_F$, where $M_{\rm Pl}$ sets the scale as in four-dimensional SUGRA. So here we shall focus on the $\mathcal{N}(U_i, S)$ factor.

To be specific, consider a Calabi-Yau-like three-fold $\mathcal{M}$ with a single ($h^{1,1}=1$) K\"ahler modulus and a relatively large $h^{2,1}$ number of complex structure moduli, so the manifold $\mathcal{M}$ has Euler number $\chi(\mathcal{M})=2(h^{1,1}-h^{2,1}) <0$.
The simplified model of interest is motivated by orientifolded orbifolds \cite{Lust:2005dy,Lust:2006zg}, given by
\begin{align}
  V &= e^{K} \left(K^{I \bar{J}} D_I W D_{\bar{J}} {\overline W} - 3\left|W \right|^2\right),\nonumber \\
  K &= K_{\rm K} + K_{\rm d} + K_{\rm cs}= -3 \ln \left(T+\bar{T} \right) -    \ln \left(S+\bar{S} \right) -  \sum_{i=1}^{h^{2,1}} \ln \left(U_i + \bar{U}_i  \right),\nonumber \\
  W &=  W_0(U_i,S) +  \cdots , \nonumber \\
 W_0(U_i,S) &= c_1 + \sum_{i=1}^{h^{2,1}} b_i U_i - S \left(c_2 + \sum_{i=1}^{h^{2,1}} d_i U_i\right) 
   +\sum_{i,j}^{h^{2,1}} \alpha_{ij}U_iU_j\;.
 \label{LVS}
\end{align}
So the contribution of $U_i$ and $S$ in $K$ gives
\begin{equation}
  \mathcal{N}(U_i, S) =e^{K_{\rm d} +K_{\rm cs}} = \frac{1}{(S +\bar{S}) \prod_i (U_i +\bar{U_i})}\;.
  \label{eq:westphal potential}
\end{equation}
The flux contribution to $W (U_i,S)$ depends on the dilation $S$ and the $h^{2,1}$ complex structure moduli $U_i$ ($i=1,2,..., h^{2,1}$), while other terms in $W$ are assumed to be independent of $U_i$ and neglected here.  We also ignore the K\"ahler uplift  term as that is also a correction. Here the parameters $c_i, b_i$,  $d_i$ are flux parameters that describes the orbifold and may be treated as independent random variables with smooth probability distributions that allow the zero values. we are interested in the physical $\Lambda$ (instead of, say, the bare $\Lambda$), so the model should include all appropriate radiative corrections.
Some explanations and justifications of the simplifications and approximations made can be found in ref.\cite{Rummel:2011cd,Sumitomo:2012vx}.

Before introducing the non-perturbative term for $T$ stabilization, the $\alpha'^3$-correction $\xi$-term for K\"ahler uplift, and the $\mathcal{D}$-term,
supersymmetric solutions are obtained with $D_JW_0=\partial_JW_0 + (\partial_JK)W_0=0$ for each $J$ where 
\begin{align}
D_S W_0 &= -c_2-\sum d_iU_i - \frac{1}{S+\bar{S}}W_0\;, \nonumber \\
D_i W_0 & = b_i -S d_i  + 2\sum_j \alpha_{ij}U_j - \frac{1}{U_i + \bar{U}_i}W_0 \;,
\label{DWsu}
\end{align}
where $i=1,2,\dots,h^{2,1}$.
Let $S=s+i \nu_0$ and $U_j = u_j +i \nu_j$. For fixed flux values $b_j, c_j$, $d_j$ and $\alpha_{ij}$, which we take real values to simplify the analysis, we first solve for $D_JW_0=0$ to determine $u_i, s$ in terms of the flux values to yield $W_0= \omega_0 (b_j, c_j, d_j, \alpha_{ij}, s, u_i)= \omega_0 (b_j, c_j, d_j, \alpha_{ij})$ and insert this into $V$ (\ref{LVS}) to solve for $T$.

To simplify, let all real flux values be fixed, so $D_JW_0=0$ immediately give,
 \begin{align}
 \label{mc4a}
    v &\equiv  vf_1+2 r_1u_i=  vf_1+2 r_2u_2=  \cdots   = vf_n +2r_nu_n \;, \nonumber \\
   f_i&=(b_i-sd_i)u_i/v, \quad r_i= \sum_j \alpha_{ij} u_j \;, \nonumber \\
     \nu_j&=0\;,
  \end{align}
and the $u_i$ are solved in terms of $s$ and one of them, say $u_1$, or equivalently, $v$. The non-geometric terms  $\alpha_{ij}=\alpha_{ji}$ may be ignored for our purpose here. Going back to eq.\eqref{DWsu} allows us to solve for $v$ and $s$ in terms of the fluxes, and
\begin{align}
W_0 &= \left.W\right|_{\rm sol}= -2\left(sc_2 +\sum_i v(p_i-f_i)\right) = 2v\;,  \nonumber \\
 p_i&=(b_i+sd_i)u_i/v, \quad p=\sum p_i\;.
\end{align} 
Scanning through the flux parameters $(c_i, b_i, d_i)$, one finds that likely ranges are
$2s \sim 2u_i \sim 5$ to $10$, for $h^{2,1} \ge 3$ and tend to $2u_i \simeq 11$ as $h^{2,1}$ becomes large. In fact, taking a large $h^{2,1}$ can provide a suppression of $\Lambda$ \cite{Sumitomo:2012vx}.
In simple orientifolds $h^{2,1} =3$. Here, we assume $h^{2,1} \gtrsim 3$. Note that $g_s = 1/{\rm Re} {(S)} \lesssim 1$, which is close to the electroweak gauge couplings and justifies the perturbative approximation implicit in our model.  

In general, we expect the order of magnitude for $V$ to be $M_S^4$.
Since we choose the Planck scale $M_{\rm Pl}=1$, 
we expect an overall factor  
\begin{equation}
\mathcal{N}(U_i, S)  \sim 10^{-4} \quad  {\rm to} \quad 10^{-3}\;,
\end{equation}
with rather large uncertainties. Inserting back the $\alpha_{ij}$ interaction terms does not change the qualitative properties of the analysis \cite{Danielsson:2012by,Blaback:2013ht,Tye:2016jzi}.
Since we have chosen the dimensionless $U_i, S$, the flux parameters $(b_i, c_i ,d_i)$ have mass dimension 3. The constraint on $W$ suggests that they have values of $m^3_{\rm EW}$ or smaller in order for a dS solution to exist.

\section{Statistical Analysis}
\label{appendix:stat}

The analysis follows that in ref.\cite{Sumitomo:2013vla}.
We consider the case $n_D=n_S=n=2$ as justified in the main text. General $n_i$ would not change the statiscial property. Due to the fact that upper bound for $\tilde{D}$ in eq.\eqref{eq:upD2}, one can simply let
\begin{equation}
	\tilde{D}=\frac{9}{10} x e^{-x}\frac{\beta-1}{3\beta}\cdot q\;,
\end{equation}
where $q\leq 1$ is a real parameter. In principle one can also randomize $q$ with some natural choice of probability distribution. Here we simply treat it as a fixed parameter. After making this assumption, we change the degree of freedom $\tilde{D}$ to paremeter $q$. Now every quantity could be expressed as functions of $(x,\beta,q)$.  The factors $A, a, \mathcal{N}, \xi$ contribute to $\Lambda$ as overall factors in $\Lambda$, so scanning them with smooth probability distribution has little impact on $P(\Lambda)$. Let us ignore them for the moment; so $\Lambda$ (\ref{eq:hL}) takes the form
\begin{equation}
	\Lambda(x,\beta,q)=\frac{2^{13/2}}{3^5}\frac{(\beta-1)^2}{\beta^2}x^{5/2}e^{-2x}\left(1-\frac{3}{5}q\right)\left(1+\frac{3}{4}q\right)\;.
\end{equation}
As long as $-4/3<q\leq1$, $\Lambda$ is positive. If one replace $A$ with $Bz$, the expression for $\Lambda$ would be a little different from this; the statistical property is the same when $\beta\gtrsim 1$. To investigate the probability distribution for small $\Lambda>0$, we simply let $q=0$. This parameter $q$ has little effect on the statistical property of $\Lambda$. With eq.\eqref{eq:cz}, we can replace the output value $x$ by the input parameter $z$ as
\begin{equation}
	x\simeq\frac{-\ln{\kappa}}{\beta-1}\;,\quad\kappa\equiv\frac{-z}{\beta^3}\;.
\end{equation}
so we express the cosmological constant as a function of the parameter $z$, or $\kappa$ of the model,
\begin{equation}
	\hat{\Lambda}(z)=\frac{2^{13/2}}{3^5}\frac{\kappa^{\frac{2}{\beta-1}}}{\beta^2\sqrt{\beta-1}}\left(-\ln{\kappa}\right)^{5/2}\;,
\end{equation}
The probability density function of $\Lambda$ could be obtained by sweeping through all parameter space. Let us first integrate out $z$ and deal with $\beta$ next. 
\begin{equation}
	P(\Lambda;\beta)=\int {\rm d} z P(z)\delta\left(\Lambda-\hat{\Lambda}(z)\right)\;.
\end{equation}
If we choose the probability distribution of $z$ is flat, $P(z)=1$ for $-1\leq z\leq0$, the above integral can be calculated easily as
\begin{align}
P(\Lambda;\beta)&=\int_{-1}^0 {\rm d} z \sum_i \frac{\delta(z-z_i)}{\left|\hat{\Lambda}'(z_i)\right|}\nonumber\\
&=\frac{3^5}{2^{11/2}}\frac{\beta^5(\beta-1)^{3/2}\kappa_0^{\frac{3-\beta}{1-\beta}}}{(-\ln{\kappa_0})^{3/2}\left(5-5\beta-4\ln{\kappa_0}\right)}\;,
\label{eq:pk}
\end{align}
where $\hat{\Lambda}(z_i)=\Lambda$ in the first line and $\kappa_0$ in the second line satisfies
\[\kappa_0^{\frac{4}{5(\beta-1)}}\ln{\kappa_0^{\frac{4}{5(\beta-1)}}}=-\frac{3^2\cdot\Lambda^{2/5}\beta^{4/5}}{2^{3/5}\cdot5\cdot(\beta-1)^{4/5}}\;.\]
Then we express eq.\eqref{eq:pk} in $\Lambda$ and this requires the solution to above equation, which involves the {\it Lambert} $\mathfrak{W}$-function, where $\mathfrak{W}(X)$ is a solution of $\mathfrak{W}e^{\mathfrak{W}}=X$,
\begin{equation}
	\ln{\kappa_0}=\frac{5(\beta-1)}{4}\mathfrak{W}_{-1}\left(-\frac{3^2\cdot\Lambda^{2/5}\beta^{4/5}}{2^{3/5}\cdot5\cdot(\beta-1)^{4/5}}\right)\;.
\end{equation}
Here, $\Lambda$ can be very small, so $\ln{\kappa_0}<0$; for $\mathfrak{W}\le -1$, we choose $\mathfrak{W}_{-1}$ as the solution. Then,
\begin{equation}
	P(\Lambda;\beta)=2^{-5/2}\cdot 3^{5}\cdot 5^{-5/2}\cdot \frac{\beta^5}{\beta-1} \frac{e^{-\frac{5(3-\beta)}{4}\mathfrak{W}_{-1}}}{\left(-\mathfrak{W}_{-1}\right)^{3/2}\left(-\mathfrak{W}_{-1}-1\right)}\;.
\end{equation}
Let us focus on the divergence behavior when $\Lambda\sim0^{+}$. So we expand the solution $\mathfrak{W}_{-1}(X)$ for small $X<0$,
\begin{equation}
\label{w-1}
\mathfrak{W}_{-1}(X)\simeq\ln{(-X)}-\ln{\left[-\ln{(-X)}\right]}+\cdots\;.
\end{equation}
Therefore, the probability ditribution for small $\Lambda$ is \footnote{keeping the next to leading term in eq.(\ref{w-1}) cuts the power of $(-\ln{\Lambda})$ compared to that in ref.\cite{Sumitomo:2013vla}.}
\begin{equation}
	P(\Lambda;\beta)\overset{\Lambda\sim0^+}{\simeq}2^{1-2\beta}\cdot 3^{\frac{5(\beta-1)}{2}}\cdot \frac{\beta^{\beta+2}}{(\beta-1)^{\beta-2}}\frac{1}{\Lambda^{\frac{3-\beta}{2}}(-\ln{\Lambda})^{\frac{5(\beta-1)}{4}}}\;.
\end{equation}
Here $ P(\Lambda;\beta)$ diverges at $\Lambda =0^+$. Notice that $\beta$ slightly larger than 1, $\beta\gtrsim 1$, so $(3-\beta)/2<1$; that is, $P(\Lambda;\beta)$ and $P(\Lambda)$ can be properly normalized, i.e., $\int {\rm d}\Lambda P(\Lambda) =1$. We see that the divergence behavior of $P(\Lambda;\beta)$ is very sensitive to the value of $\beta$, while the dependence of $P(\Lambda;\beta)$ on the other parameters $(a,A,\xi,\mathcal{N},q)$  and $P(z)$ are much less sensitive.

To obtain $P(\Lambda)$, we have to scan over values of $\beta$. Recall that $a=2\pi/N_1$ and $b=2\pi/N_2$ for the two gauge groups $SU(N_1)$ and $SU(N_2)$, so $\beta=b/a=N_1/N_2 >1$, where $N_1>N_2$ by convention. We should scan over $N_1=3,4,\dots,N_\text{max}$ and $N_2=2,3,\dots,N_1-1$. This happens when $\beta>1$ is smallest, and for $N_\text{max}  \gg 2$, we have
$$\beta_{\rm min} =\frac{N_\text{max}}{N_\text{max} -1}\; \to\; \frac{3-\beta_{\rm min}}{2}= 1 -\frac{1}{2(N_{\rm max}-1)}\;.$$
The most dominant distribution at $\Lambda \gtrsim 0$, which diverges at $\Lambda \simeq 0^{+}$, is the one with $\beta_{\rm min}$. So, to a good approximation we may simply set $P(\Lambda)\simeq P(\Lambda;\beta_{\rm min})$, 
\begin{equation}
P(\Lambda)\overset{\Lambda \sim 0^{+}}{\simeq} \frac{N_{\rm max}^3}{2(N_{\rm max}-1)^4} \frac{\Lambda^{-1+\frac{1}{2N_{\rm max}}}}{(-\ln \Lambda)^{5/[4(N_{\rm max}-1)]}} \sim \frac{1}{2N_{\rm max}}\Lambda^{-1+\frac{1}{2N_{\rm max}}}\;,
\label{Lpower}
\end{equation}
where we have dropped the logarithmic factor, since the logarithmic divergence is very weak compared to the power divergence, and the power of the logarithmic factor here is very small. If we naively extend this to all $\Lambda >0$, $\int_0^{\infty} P(\Lambda) \, {\rm d}\Lambda$ diverges. In the text, we normalize $P(\Lambda)$ (\ref{Papprox}) or (\ref{Lpower}) by setting $P(\Lambda)=0$ for $\Lambda>1$. 

We find it convenient to use an alternative way to cut-off, for example, to fit $P(\Lambda)$ with the Weibull distribution (because it is simple), for $\Lambda  \ge 0$, 
\begin{equation}
P(\Lambda)\approx \tilde{P}(\Lambda)=k\Lambda^{k-1} e^{-\Lambda^k}, \quad \quad k=\frac{1}{2N_{\rm max}}\;,
\end{equation}
where $\int_0^{\infty} \tilde{P}(\Lambda) \, {\rm d}\Lambda =1$.  This is a prime example of what we have in mind for a typical monotonic probability distribution for $P(\Lambda)$. From
\[
\int_0^{\Lambda_{Y \%}}\tilde{P}(\Lambda) {\rm d}\Lambda=Y \%\;,
\]
it is easy to obtain
$$\Lambda_{Y \%}= \left[- \ln\{1-(Y/100)\}\right]^{1/k}\;.$$ 
So matching $\Lambda_{Y\%}$ to the observed $\Lambda_{\rm obs} \sim 10^{-120}M_{\rm Pl}^4$ yields
$$N_{\rm max} =\frac{1}{2} \ln \Lambda_{\rm obs} /\ln \left[- \ln\{1-(Y/100)\}\right]\;,$$
and one finds $N_{\rm max} \simeq 380$ for $\Lambda_{50 \%} \simeq \Lambda_{\rm obs}$.

\bibliographystyle{JHEP}
\bibliography{refs}

\end{document}